\documentclass[prx,aps,twocolumn,showpacs,amsmath,amssymb,superscriptaddress,floatfix]{revtex4-2}
\usepackage{booktabs}
\usepackage{multirow}
\usepackage{graphicx}
\usepackage[export]{adjustbox}
\usepackage{dcolumn}
\usepackage{upgreek}
\usepackage{bm,dsfont}
\usepackage{amssymb,mathtools}
\usepackage{booktabs}
\usepackage{color}
\usepackage{microtype}
\usepackage[T1]{fontenc}
\usepackage[shortlabels]{enumitem}
\usepackage{physics}
\usepackage{derivative}
\usepackage{pgfplots}
\usepackage{placeins}
\usepackage{sidecap}
\usepackage{orcidlink}
\usepackage{longtable}
\usepackage[linesnumbered,ruled,vlined]{algorithm2e}
\allowdisplaybreaks
\hypersetup{
colorlinks = true,
linkcolor = [rgb]{0.87,0.18,0.22},
citecolor = [rgb]{0.0,0.60,0.33},
urlcolor = [rgb]{0.20, 0.30, 0.54}}

\definecolor{eggplant}{RGB}{180,33,147}
\makeatletter\renewcommand{\fnum@table}[1]{\tablename~\thetable.}\makeatother

\definecolor{citecolor}{rgb}{0.0,0.60,0.33}

\definecolor{mygreen}{HTML}{299D8F}
\definecolor{myblue}{HTML}{407BD0}
\definecolor{myred}{HTML}{D87659}

\definecolor{racolor}{HTML}{306B34}
\definecolor{rbcolor}{HTML}{6594B8}
\definecolor{authorcolor}{HTML}{592E83}

\usepackage{bbm}
\usepackage{times}

\begin{document}

\title{Learning Thermoelectric Transport from Crystal Structures via Multiscale Graph Neural Network}

\author{Yuxuan Zeng\orcidlink{0009-0004-8954-4377}}
\affiliation{School of Integrated Circuits, Wuhan University, Wuhan 430072, PR China}

\author{Wei Cao\orcidlink{0000-0001-5131-0728}
}
\email{wei\_cao@whu.edu.cn}
\affiliation{School of Integrated Circuits, Wuhan University, Wuhan 430072, PR China}
\affiliation{Key Laboratory of Artiﬁcial Micro- and Nano-Structures of Ministry of Education, School of Physics and Technology, Wuhan University, Wuhan 430072, PR China}

\author{Yijing Zuo}
\affiliation{Key Laboratory of Artiﬁcial Micro- and Nano-Structures of Ministry of Education, School of Physics and Technology, Wuhan University, Wuhan 430072, PR China}

\author{Fang Lyu}
\affiliation{Key Laboratory of Artiﬁcial Micro- and Nano-Structures of Ministry of Education, School of Physics and Technology, Wuhan University, Wuhan 430072, PR China}

\author{Wenhao Xie}
\affiliation{School of Integrated Circuits, Wuhan University, Wuhan 430072, PR China}

\author{Tan Peng}
\affiliation{Key Laboratory of Artiﬁcial Micro- and Nano-Structures of Ministry of Education, School of Physics and Technology, Wuhan University, Wuhan 430072, PR China}

\author{Yue Hou}
\affiliation{School of Integrated Circuits, Wuhan University, Wuhan 430072, PR China}

\author{Ling Miao}
\affiliation{School of Optical and Electronic Information, Huazhong University of Science and Technology, Wuhan 430072, PR China}

\author{Ziyu Wang\orcidlink{0000-0001-9718-1263}
}
\email{zywang@whu.edu.cn}
\affiliation{School of Integrated Circuits, Wuhan University, Wuhan 430072, PR China}
\affiliation{Key Laboratory of Artiﬁcial Micro- and Nano-Structures of Ministry of Education, School of Physics and Technology, Wuhan University, Wuhan 430072, PR China}
\affiliation{School of Physics and Microelectronics, Key Laboratory of Materials Physics of Ministry of Education, Zhengzhou University, Zhengzhou 450001, PR China}

\author{Jing Shi}
\affiliation{Key Laboratory of Artiﬁcial Micro- and Nano-Structures of Ministry of Education, School of Physics and Technology, Wuhan University, Wuhan 430072, PR China}

\date{\today}
\begin{abstract}
Graph neural networks (GNNs) are designed to extract latent patterns from graph-structured data, making them particularly well suited for crystal representation learning. Here, we propose a GNN model tailored for estimating electronic transport coefficients in inorganic thermoelectric crystals. The model encodes crystal structures and physicochemical properties in a multiscale manner, encompassing global, atomic, bond, and angular levels. It achieves state-of-the-art performance on benchmark datasets with remarkable extrapolative capability. By combining the proposed GNN with \textit{ab initio} calculations, we successfully identify compounds exhibiting outstanding electronic transport properties and further perform interpretability analyses from both global and atomic perspectives, tracing the origins of their distinct transport behaviors. Interestingly, the decision process of the model naturally reveals underlying physical patterns, offering new insights into computer-assisted materials design.
\end{abstract}

\maketitle

\section{Introduction}
\label{sec:intro}

\underline{T}hermo\underline{e}lectric (TE) materials convert temperature gradients into electrical potential differences by driving carrier transport via the Seebeck effect~\cite{he2017advances}. This process involves no mechanical vibrations, produces no pollutant emissions, and is easy to integrate, making TE materials widely applicable in flexible and wearable electronic devices~\cite{sun2025modular,qijian2,kraemer2016concentrating}. Evaluation of TE performance requires a comprehensive consideration of the Seebeck coefficient ($S$), electrical conductivity ($\sigma$), and thermal conductivity ($\kappa$). These properties are collectively quantified by the dimensionless figure of merit, $zT = \frac{S^2 \sigma T}{\kappa}$. The power factor, defined as ${\rm PF} = S^2 \sigma$, can also serve as an independent metric to assess the electrical transport performance of a material. The thermal transport in TE materials involves the combined contributions of both electronic and phononic carriers, i.e., $\kappa = \kappa_{\rm e} + \kappa_{\rm ph}$. The corresponding electronic and phonon scattering mechanisms vary under different temperature and carrier conditions, making both $\kappa_{\rm e}$ and $\kappa_{\rm ph}$ crucial for the overall thermal conductivity performance~\cite{kim2016electronic}. Given the complexity of TE transport mechanisms and the nearly infinite combinations of elements and structures, traditional trial-and-error experimentation and \textit{ab initio} computations, aimed at discovering potential outstanding TE materials, face significant challenges like high costs and long development cycles. Apparently, there is an increasing need for more efficient computational models and strategies to accelerate the identification of promising TE materials.

To accelerate the high-throughput screening, artificial intelligence has emerged as a promising paradigm for exploring potential candidates~\cite{PRE,PRXEnergy.3.011002}. While machine learning techniques have been increasingly adopted to predict the properties of TE materials, accurately representing crystal structures remains a major challenge. Existing approaches either rely solely on chemical formulas~\cite{Antunes2023,XU2021110625,al2024high} or employ empirical descriptors~\cite{yan2015material,toriyama2023material,huang2023exploring}. Features derived purely from chemical formulas have the advantage of being easy to construct, but they inherently overlook the influence of crystal topology on material properties. A representative example is diamond and graphite, which share the same chemical formula ``C'' but exhibit vastly different properties, e.g., diamond is an insulator, while graphite is a good conductor. Since lattice distortion serves as an important strategy in tuning TE transport~\cite{cw-prb,lf-zhengfu}, structural representations are therefore indispensable. On the other hand, constructing empirical descriptors often requires resource-intensive \textit{ab initio} calculations or experimental measurements.

To address above issues, an effective approach is to employ graph neural networks (GNNs) to model crystal structures for predicting their physicochemical properties. GNNs were first applied to molecules~\cite{scarselli2008graph,wang2022molecular,coley2017convolutional}, owing to their discrete structures and accessible descriptors (e.g., SMILES~\cite{m2015application}). In contrast, crystal modeling is more complicated due to structural periodicity and symmetry. CGCNN~\cite{cgcnn} tackled this challenge by applying graph convolutions to iteratively update atomic embeddings using neighbor and bond information. Building on this idea, a series of sophisticated derivatives have continuously improved predictive performance through clever architectural designs, consistently pushing the benchmark~\cite{choubisa2023interpretable,alignn,karamad2020orbital,park2020developing}. While such models achieve \textit{ab initio}-level accuracy for general properties like formation energy ($E_{\rm f}$), band gap ($E_{\rm g}$), and bulk modulus, they often underperform in TE transport prediction~\cite{cralinet}, sometimes even lagging behind formula-based models~\cite{Antunes2023}. This illustrates the \textit{No Free Lunch} theorem~\cite{nfl}, highlighting the necessity for task-specific designs. For instance, in predicting adsorption performance, simultaneously considering local structural attributes and global textural properties helps achieve state-of-the-art (SOTA) performance~\cite{lin2025unified}. Likewise, the piezoelectric tensor depends on the choice of coordinate system and crystal symmetry, making it necessary to employ equivariant models that capture the relationship between crystal symmetry and the target tensor~\cite{dong2025accurate}. Therefore, learning TE transport also requires incorporating physics-guided design, rather than simply applying standard baseline models.

In this work, we propose a GNN-guided pipeline for high-throughput estimation of TE properties in inorganic crystalline materials. Beyond the commonly adopted atomic, bond, and angular features, we further incorporate global crystal-level descriptors to capture overarching physicochemical characteristics of the structure. Trained on a dataset~\cite{ricci2017ab} calculated by density functional theory (DFT), our model achieves SOTA accuracy in predicting TE transport properties. To enhance interpretability, we conduct sensitivity analysis to quantify the contribution of each input feature to the model output. Notably, this pipeline requires no domain-specific prior knowledge and offers a general and efficient tool to accelerate the discovery of promising TE materials.

\section{Results}

\subsection{Crystal graph representation and network architecture design}
\label{sec:graph-net-arch}

\begin{figure}[ht]
    \centering
    \includegraphics[width=\columnwidth]{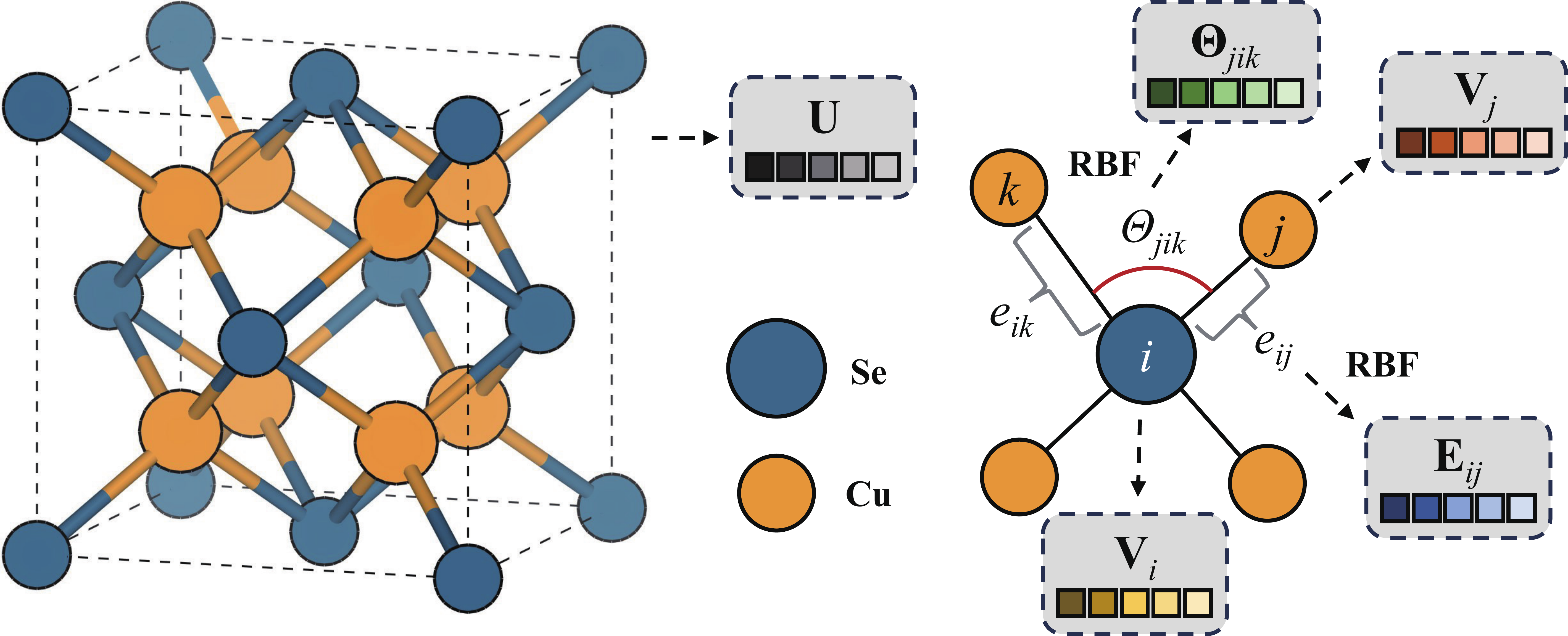}
    \caption{Graph representation of crystal structures. Taking CuSe$_2$ as an illustration, the crystal structure can be mathematically represented by a set of hierarchical vectorial descriptors at multiple scales, corresponding respectively to the global statistical properties, atomic sites, chemical bonds, and bond angles.}
    \label{fig:crys-graph}
\end{figure}

Graph, as a non-Euclidean data structure, is well-suited for representing complex topologies. An undirected graph $\mathcal{G}$ consists of node vectors $\mathbf{V}$ and edge vectors $\mathbf{E}$, i.e., $\mathcal{G} = \left(\mathbf{V}, \mathbf{E} \right)$~\cite{wu2020comprehensive}. In terms of fundamental properties alone, representing crystal information using $(\mathbf{V}, \mathbf{E})$ is sufficient to achieve DFT-level accuracy. However, complex properties, including TE transport, are sensitive to structural variations such as bond angles and local geometric distortions. Therefore, introducing bond angles $\mathbf{\Theta}$ as an additional scale for crystal representation~\cite{alignn} is reasonable. In the specific context of TE transport, existing works indicate that global physicochemical properties of the crystal and their statistics (denoted as $\mathbf{U}$) can achieve reliable accuracy and possess interpretable physical meaning~\cite{vipin2024machine,choubisa2023closed}. As noted in Sec.~\ref{sec:intro}, these descriptors are generated based on chemical formulas and lack structural representation. Our approach couples the formula-based descriptors $\mathbf{U}$ with the structure-based $(\mathbf{V}, \mathbf{E}, \mathbf{\Theta})$, enabling multi-scale modeling of crystals, see Fig.~\ref{fig:crys-graph}.

In GNNs, a distinctive operation known as ``graph convolution'' ~\cite{pmlr-v48-niepert16} is employed, which is similar to but also distinct from convolutions in convolutional neural networks (CNNs). In CNNs, convolutions aim to capture local patterns in images, such as edges, corners, and textures. Similarly, graph convolution in GNNs enables each node to aggregate information from its neighboring nodes, thereby refining its own embedding. In our case, we adopt a periodic $k$-nearest neighbors ($k$-NN) scheme, in which each atom in the primitive cell updates its embedding based on the $k$ closest atoms and the corresponding edge information within the periodic lattice. In practice, we set $k=8$; smaller values may lead to information loss, whereas larger values significantly increase computational cost. It is important to note that $k=8$ does not limit the receptive field of an atom. Through multiple rounds of convolution, structural information from more distant atoms is also propagated to each node, albeit with diminishing influence, which aligns well with physical intuition~\cite{zeier2016thinking}.
\begin{figure*}[t]
    \centering
    \includegraphics[width=\textwidth]{net.pdf}
    \caption{The proposed GNN framework for crystal structure representation learning. (a) The overall architecture of the TECSA-GNN. Distinct colours are employed to indicate feature vectors at different scales. The superscript ${(0)}$ designates the embedding representation subsequent to dimensional alignment by means of an MLP projection, whereas ``$N$'' corresponds to the embedding yielded upon passage through the $N$-th graph convolutional module (GConv \#$N$). (b) Detailed schematic of an individual graph convolutional module. In this representation, both ``Concatenation'' and ``$\oplus$'' signify the concatenation of embeddings; ``$\sum$'' denotes element-wise summation realised via residual connections; and ``$\otimes$'' indicates element-wise multiplication.}
    \label{fig:net}
\end{figure*}

As illustrated in Fig.~\ref{fig:net}, we refer to the proposed architecture as the \underline{T}hermo\underline{e}lectric \underline{C}rystal \underline{S}tructure-\underline{A}ware Graph Neural Network (TECSA-GNN). The model represents a crystal as a multigraph,
\begin{align}
\mathcal{G}=\left\{\mathbf{U},\mathbf{V}_i,\mathbf{E}_{ij},\boldsymbol{\Theta}_{jik}\right\},
\label{eq:graph-define}
\end{align}
where $\mathbf{V}_i$ denotes atomic features, $\mathbf{E}_{ij}$ bond features, $\boldsymbol{\Theta}_{jik}$ angular features centered at atom $i$, and $\mathbf{U}$ global crystal-level descriptors. Bond lengths and bond angles are expanded using radial basis functions (RBF)~\cite{alignn,acosta1995radial}.

To enable coherent convolution across heterogeneous feature types, we first align their dimensionalities by projecting atomic, bond, and angular features into a shared latent space using independent MLPs. Each MLP performs a linear transformation followed by layer normalization $\mathcal{N}\left(\cdot\right)$ and nonlinear activation $\sigma\left(\cdot\right)$,
\begin{align}
\begin{cases}
\tilde{\mathbf{V}}_i=\sigma\left(\mathcal{N}\left(W_x\mathbf{V}_i+b_x\right)\right),\\
\tilde{\mathbf{E}}_{ij}=\sigma\left(\mathcal{N}\left(W_e\mathbf{E}_{ij}+b_e\right)\right),\\
\tilde{\boldsymbol{\Theta}}_{jik}=\sigma\left(\mathcal{N}\left(W_\theta\boldsymbol{\Theta}_{jik}+b_\theta\right)\right),
\end{cases}
\end{align}
thereby ensuring that all feature types reside in a common embedding space before message passing. Here $\sigma(\cdot)$ denotes the ReLU activation, and
\begin{align}
\mathcal{N}(\mathbf{v})=\boldsymbol{\gamma}\odot\frac{\mathbf{v}-\langle\mathbf{v}\rangle}{\sqrt{\mathrm{Var}(\mathbf{v})+\varepsilon}}+\boldsymbol{\beta}
\end{align}
denotes layer normalization. The initial embedding and its evolution across $N$ graph convolution layers are defined as
\begin{align}
\mathbf{H}^{(0)}=\left\{\tilde{\mathbf{V}}_i,\tilde{\mathbf{E}}_{ij},\tilde{\boldsymbol{\Theta}}_{jik}\right\},\quad \mathbf{H}^{(N)}=\Phi^{(N)}\left(\mathbf{H}^{(N-1)}\right).
\end{align}

Graph-level representations are obtained via mean pooling,
\begin{align}
\bar{\mathbf{V}}=\frac{1}{|\mathcal{V}|}\sum_{i\in\mathcal{V}}\mathbf{V}_i^{(N)},\quad \bar{\mathbf{E}}=\frac{1}{|\mathcal{E}|}\sum_{(i,j)\in\mathcal{E}}\mathbf{E}_{ij}^{(N)},
\end{align}
and the pooled embeddings are concatenated with the global descriptor $\mathbf{U}$ to form
\begin{align}
\mathbf{h}_{\mathcal{G}}=\left[\left(W_n\bar{\mathbf{V}}+c_n\right)^\top\oplus\left(W_e\bar{\mathbf{E}}+c_e\right)^\top\oplus\mathbf{U}^\top\right]^\top,
\end{align}
which is mapped to the transport-coefficient prediction, $\hat{y}=f_\theta\left(\mathcal{N}\left(\mathbf{h}_{\mathcal{G}}\right)\right)\in\mathbb{R}^{1}$. Hierarchical message passing proceeds in the order $\boldsymbol{\Theta}\rightarrow\mathbf{E}\rightarrow\mathbf{V}$. In the $l$-th graph convolution layer, node and edge embeddings are updated simultaneously,
\begin{align}
\begin{cases}
\mathbf{X}^{(l)}=\Phi^{(l)}\left(\mathbf{X}^{(l-1)},\mathbf{E}^{(l-1)}\right),\\
\mathbf{E}^{(l)}=\Psi^{(l)}\left(\mathbf{E}^{(l-1)},\boldsymbol{\Theta}\right),
\end{cases}
\end{align}
and the edge message is constructed as
\begin{align}
\mathbf{m}_{i\leftarrow j}^{(l)}=\sigma\left[\mathcal{N}\left(\mathbf{X}_i^{(l-1)}\oplus\mathbf{X}_j^{(l-1)}\oplus\mathbf{E}_{ij}^{(l-1)}\right)\right].
\end{align}

The attention coefficient is defined by $\alpha_{ij}^{(l)}=\mathbf{w}^{\top}\mathbf{m}_{i\leftarrow j}^{(l)}$. Node embeddings are updated via normalized attention aggregation,
\begin{align}
\mathbf{X}_i^{(l)}=\mathcal{N}\left(\mathbf{X}_i^{(l-1)}+\sum_{j\in\mathcal{N}(i)}\frac{\exp\left(\alpha_{ij}^{(l)}\right)}{\sum_{k\in\mathcal{N}(i)}\exp\left(\alpha_{ik}^{(l)}\right)}\mathbf{m}_{i\leftarrow j}^{(l)}\right).
\end{align}

To model angular coupling, message passing is performed on the line graph. For adjacent bonds $(i,j)$ and $(i,k)$ sharing atom $i$, the angular message is defined as
\begin{align}
\mathbf{r}_{jik}^{(l)}=\sigma\left[\mathcal{N}\left(\mathbf{E}_{ij}^{(l-1)}\oplus\mathbf{E}_{ik}^{(l-1)}\oplus\boldsymbol{\Theta}_{jik}\right)\right].
\end{align}
Bond embeddings are updated via line-graph aggregation,
\begin{align}
\mathbf{E}_{ij}^{(l)}=\mathcal{N}\left(\mathbf{E}_{ij}^{(l-1)}+\frac{1}{|\mathcal{M}(i,j)|}\sum_{k\in\mathcal{M}(i,j)}\mathbf{r}_{jik}^{(l)}\right),
\end{align}
where $\mathcal{M}(i,j)$ denotes the set of neighboring bonds forming angle triplets with bond $(i,j)$. Each atom is connected to its eight nearest neighbors to balance computational efficiency and predictive accuracy. Bonding and angular relations are extracted using the \texttt{pymatgen} toolkit~\cite{pymatgen}.

\subsection{Quantitative evaluation, unsupervised learning and fine-tuning}
\label{sec:performance}

\begin{figure*}[t]
    \centering
    \includegraphics[width=\textwidth]{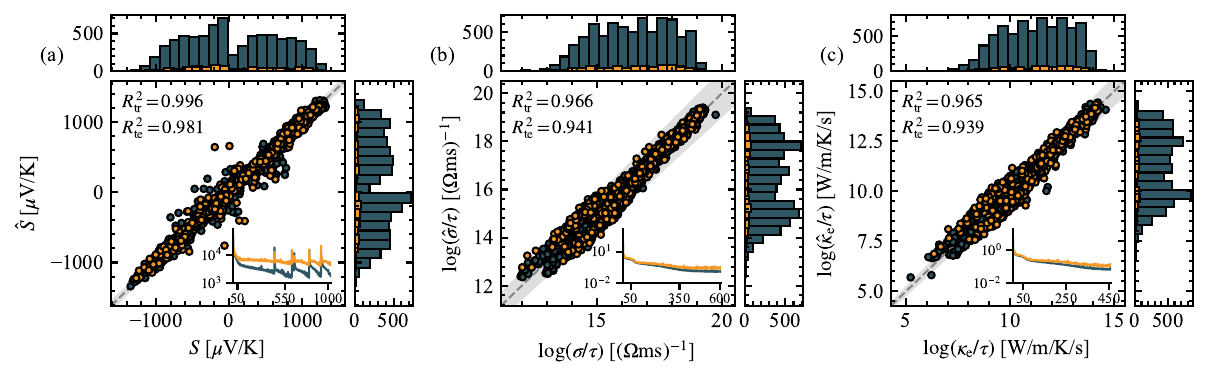}
    \caption{Error evaluation for the TECSA-GNN model. (a)-(c) The parity plots for $S$, $\log(\sigma/\tau)$, and $\log(\kappa_{\rm e}/\tau)$, where yellow and blue scatter points represent the test and training samples, respectively. The lower-right inset of each subplots shows the model loss curve, with the colors corresponding to the same sample groups as in the scatter plot.}
    \label{fig:performance}
\end{figure*}

In essence, TECSA-GNN may be regarded as a learnable function whose task is to map a crystal graph $\mathcal{G}$ onto a quantitative descriptor of TE transport, denoted $\mathcal{Y}$, i.e., $(\mathcal{G}; \omega)\rightarrow \mathcal{Y}$, where $\mathcal{Y}=\{S, \sigma/\tau, \kappa_{\rm e}/\tau\}$ represents the set of three canonical quantities employed to assess TE transport performance. The crystal graph $\mathcal{G}$ is the abstract representation defined in Eq.~\eqref{eq:graph-define}, and $\omega$ denotes the network parameters to be optimised during training. Concretely, the regression task is cast as the minimisation of a loss function, for which we adopt the mean squared error (MSE), i.e., $\mathcal{L} = \frac{1}{N_{\rm B}}\sum_{i=1}^{N_{\rm B}} \left(y_i - \hat{y}_i\right)^2$, where $\hat{y}_i$ designates the model prediction of the target value, and $N_{\rm B}$ denotes the number of samples contained in a mini-batch, treated here as a hyperparameter. 

In our case, the doping type and concentration of the material are concatenated with the global feature vector $\mathbf{U}$ as part of the input representation, while model training is conducted under distinct temperature conditions. In addition, the band gap $E_{\rm g}$, which constitutes one of the decisive parameters governing electronic transport properties, is directly provided by the dataset and likewise incorporated into $\mathbf{U}$. It must nevertheless be recognised that, for entirely unseen materials, $E_{\rm g}$ is typically unavailable. A practically effective remedy is to invoke SOTA pretrained models that infer $E_{\rm g}$ from the crystal structure \textit{ab initio}, with the estimated value $\hat{E}_{\rm g}$ serving as a surrogate. Contemporary advances in AI-based prediction of $E_{\rm g}$ already permit accuracies commensurate with density functional theory; further details are given in~\cite{supp}.

\begin{figure}[ht]
    \centering
    \includegraphics[width=\columnwidth]{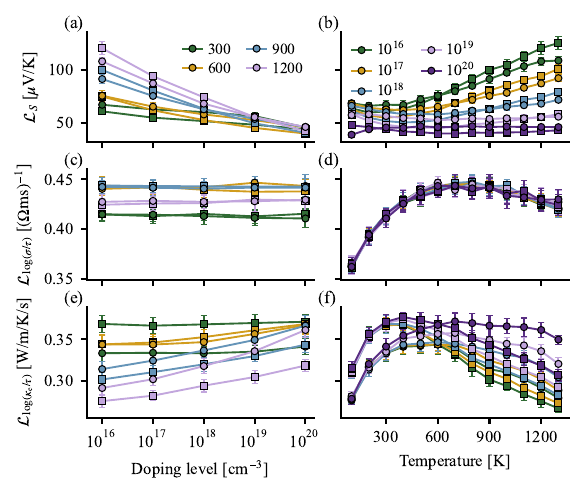}
    \caption{10-fold cross-validation performance of the TECSA-GNN. (a), (c), and (e) show the dependence of $S$, $\log(\sigma/\tau)$, and $\log(\kappa_{\rm e}/\tau)$ on carrier concentration, with $T=\left\{300,600,900,1200\right\}~{\rm K}$. (b), (d), and (f) depict the dependence of $S$, $\log(\sigma/\tau)$, and $\log(\kappa_{\rm e}/\tau)$ on temperature, with $n=\left\{10^{16},10^{17},10^{18},10^{19},10^{20}\right\}~{\rm cm}^{-3}$. Circular markers denote $n$-type samples, while square markers denote $p$-type. Error bars represent the mean and standard deviation of the MAE across the 10 folds.}
    \label{fig:pref-curve}
\end{figure}

\begin{table*}
\centering
\caption{Benchmark of test set results in terms of MAE for TE transport properties and basic material properties. Bold values indicate SOTA performance. For the carrier concentration, we adopt the standard setting of $n=10^{19}~\mathrm{cm}^{-3}$.}
\label{tab:benchmark}
\begin{tabular}{cccccccccc}
\toprule
Property                       & \multicolumn{1}{c}{Unit}                                & \multicolumn{1}{c}{Temp.}      & Type             & CrabNet~\cite{crabnet} & CGCNN~\cite{cgcnn}  & GeoCGNN~\cite{geocgnn} & ALIGNN~\cite{alignn}         & CraLiGNN~\cite{cralinet}         & TECSA-GNN       \\
\midrule
$S$                            & \multirow{2}{*}{$\mu$V/K}                               & \multirow{2}{*}{600 K}               & \textit{n}                & 89.024  & 109.11 & 97.72  & 84.267         & 74.645           & \textbf{60.686} \\
$S$                            &                                                         &                                      & \textit{p}       & 86.812  & 113.32  & 104.97  & 80.925         & 79.79            & \textbf{54.333} \\
$S$                            & \multirow{2}{*}{$\mu$V/K}                               & \multirow{2}{*}{900 K}               & \textit{n}       & 86.756  & 92.528 & 106.53 & 81.815         & 74.685           & \textbf{66.079} \\
$S$                            &                                                         &                                      & \textit{p}       & 96.165  & 102.26 & 111.37 & 88.203         & 81.209           & \textbf{66.408} \\
$\log\left(\sigma/\tau\right)$ & \multirow{2}{*}{$\log\left(\Omega{\rm ms}\right)^{-1}$} & \multirow{2}{*}{600 K}               & \textit{n}       & 0.563   & 0.586  & 0.526   & 0.493          & 0.454            & \textbf{0.265}  \\
$\log\left(\sigma/\tau\right)$ &                                                         &                                      & \textit{p}       & 0.615   & 0.568  & 0.559   & 0.497          & 0.46             & \textbf{0.293}  \\
$\log\left(\sigma/\tau\right)$ & \multirow{2}{*}{$\log\left(\Omega{\rm ms}\right)^{-1}$} & \multirow{2}{*}{900 K}               & \textit{n}       & 0.608   & 0.540  & 0.518   & 0.471          & 0.429            & \textbf{0.261}  \\
$\log\left(\sigma/\tau\right)$ &                                                         &                                      & \textit{p}       & 0.601   & 0.547  & 0.556   & 0.502          & 0.438            & \textbf{0.286}  \\
$\log\left(\kappa_{\rm e}/\tau\right)$ & \multirow{2}{*}{$\log\left({\rm W/m/K/s}\right)$} & \multirow{2}{*}{600 K} & \textit{n} & 0.634 & 0.598 & 0.563 & 0.546 & 0.476 & \textbf{0.352} \\
$\log\left(\kappa_{\rm e}/\tau\right)$ &                                                           &                                      & \textit{p} & 0.662 & 0.612 & 0.609 & 0.559 & 0.504 & \textbf{0.365} \\
$\log\left(\kappa_{\rm e}/\tau\right)$ & \multirow{2}{*}{$\log\left({\rm W/m/K/s}\right)$} & \multirow{2}{*}{900 K} & \textit{n} & 0.586 & 0.564 & 0.555 & 0.493 & 0.449 & \textbf{0.350} \\
$\log\left(\kappa_{\rm e}/\tau\right)$ &                                                           &                                      & \textit{p} & 0.603 & 0.538 & 0.582 & 0.513 & 0.449 & \textbf{0.319} \\
$E_{\rm g}$                    & \multicolumn{1}{c}{eV}                                  & \multicolumn{1}{c}{-} & - & 0.271   & 0.400  & 0.289   & \textbf{0.224} & - & 0.397           \\
$E_{\rm f}$                    & \multicolumn{1}{c}{eV/atom}                             & \multicolumn{1}{c}{-} & - & 0.081   & 0.040  & 0.030   & \textbf{0.022} & - & 0.054           \\
\botrule
\end{tabular}
\end{table*}

In practice, we model $\log(\sigma/\tau)$ and $\log(\kappa_{\rm e}/\tau)$ instead of $\sigma/\tau$ and $\kappa_{\rm e}/\tau$ themselves to alleviate skewness arising from their wide-ranging distribution~\cite{higgins2008meta}. We employed 10-fold cross-validation to evaluate the predictive performance of the model. Fig.~\ref{fig:performance} illustrates the results for a fixed data split, where the training accuracy reflects the model's learning capacity on the given dataset, while the validation accuracy quantifies its inference capability. Fig.~\ref{fig:pref-curve} further presents how variations in data distribution, temperature, doping type, and doping concentration influence the model's inference performance. Importantly, to prevent data leakage, all entries corresponding to the same material under different doping conditions were consistently assigned to the same subset. As anticipated, the prediction error for $S$ is the smallest among the three targets, while the errors for $\log(\sigma/\tau)$ and $\log(\kappa_{\rm e}/\tau)$ are comparable, largely reflecting the approximation inherent in $\tau$. Figs.~\ref{fig:performance}(d)-(f) depict the training and validation loss curves; $\mathcal{L}_{\rm Train}$ and $\mathcal{L}_{\rm Test}$ for $S$ are closely aligned, indicative of robust generalisation. Slight underfitting is observed for $\log(\sigma/\tau)$ and $\log(\kappa_{\rm e}/\tau)$, but overall accuracy remains acceptable. The principal advantage of TECSA-GNN lies in its unified architecture, which accommodates the simultaneous learning of $\{S, \sigma/\tau, \kappa_{\rm e}/\tau\}$ without necessitating task-specific adjustments. Further details on hyperparameter settings are provided in~\cite{supp}.

\begin{figure*}[t]
    \centering
    \includegraphics[width=\textwidth]{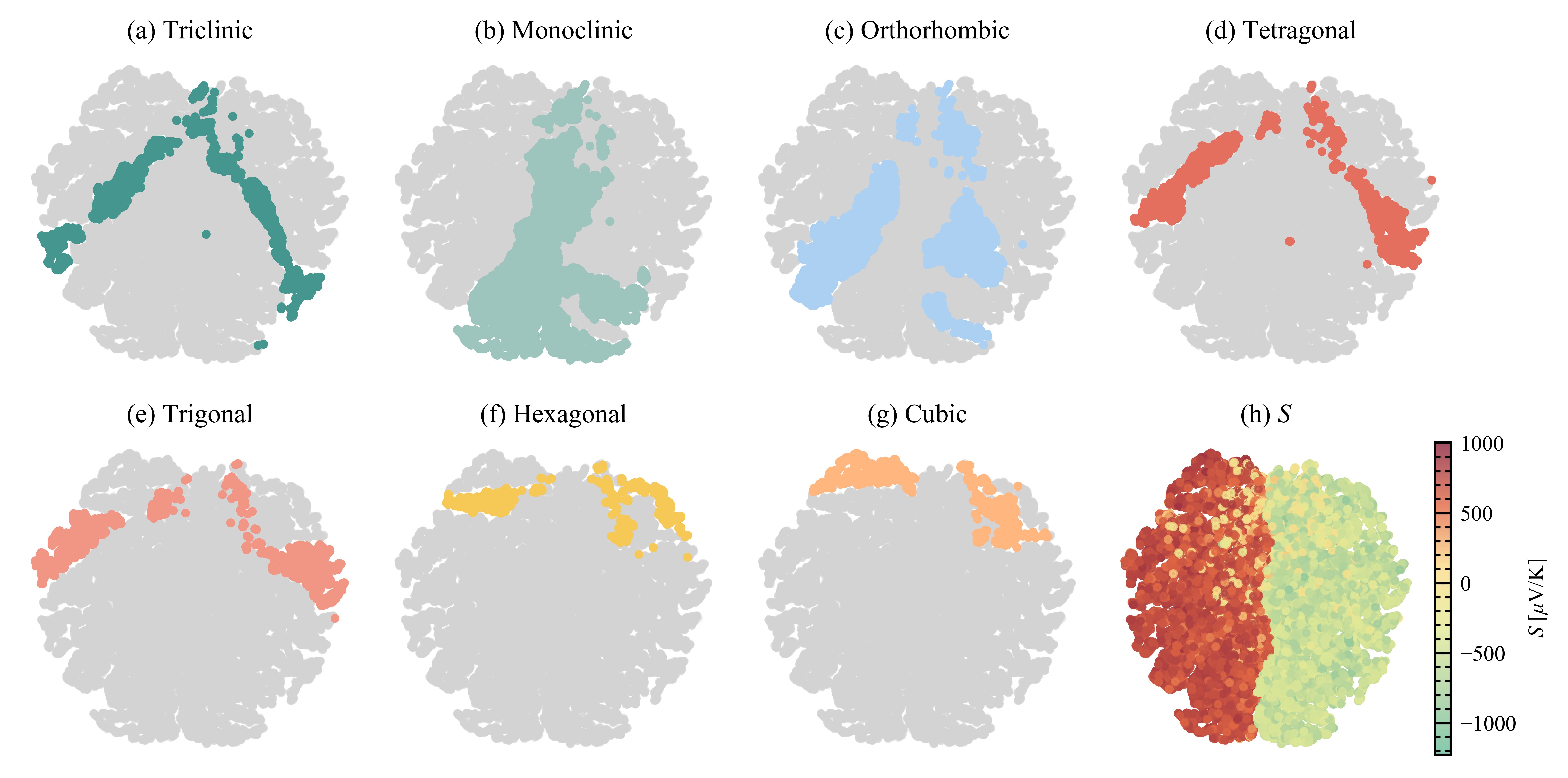}
    \caption{$t$-SNE visualization of sample embeddings across different crystal systems. (a)-(g) Each colour denotes one of the seven crystal systems in the two-dimensional $t$-SNE plane, with grey points indicating samples from other systems. (h) Colour mapping of the Seebeck coefficient $S$ illustrates both its magnitude and spatial distribution.}
    \label{fig:tsne-cs-s}
\end{figure*}

\begin{figure*}
    \centering
    \includegraphics[width=\textwidth]{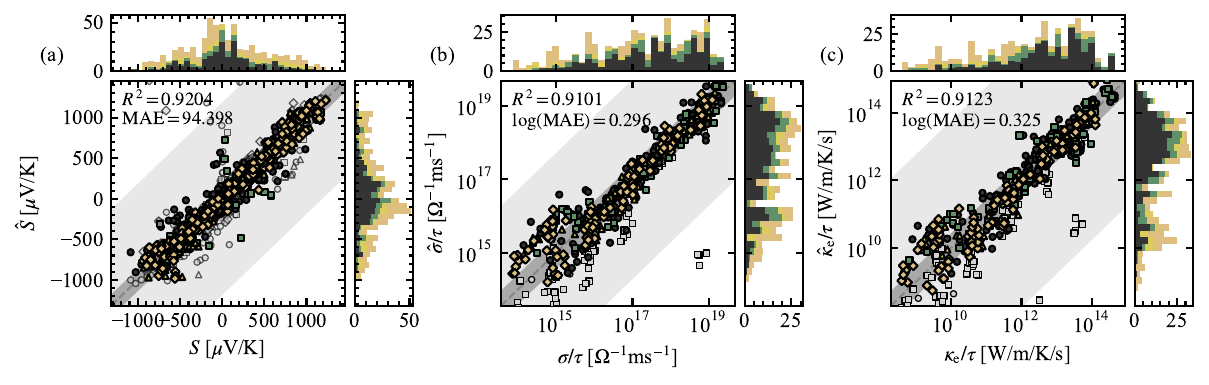}
    \caption{Parity plots of TECSA-GNN after fine-tuning on special structures. (a) $S$, (b) $\log(\sigma/\tau)$, and (c) $\log(\kappa_{\rm e}/\tau)$. Only test-set samples are shown. Circles, squares, triangles, and diamonds correspond to HH, FH, ZB, and RS structures. Grey markers denote predictions from the pretrained model, while coloured markers indicate predictions after fine-tuning. Light-grey bands represent one standard deviation of the prediction error of the pretrained model, with the corresponding range after fine-tuning shown in dark grey. Histograms along the top and right of each panel display the distributions of DFT reference values and model estimates, respectively.}
    \label{fig:fine-tune}
\end{figure*}

We provide a comprehensive benchmark to assess the accuracy of different models in estimating TE transport properties, as summarised in Tab.~\ref{tab:benchmark}. Among the models included for comparison, only CrabNet~\cite{crabnet} characterises mateirals purely in terms of their chemical compositions, whereas all others are GNNs that exploit structural descriptors. In this benchmark, we follow the convention of Ref.~\cite{cralinet} by setting the carrier concentration to $n=10^{19}~\mathrm{cm}^{-3}$ and the temperatures to 600~K and 900~K, with separate evaluations performed for $n$-type and $p$-type doping. As compared in Tab.~\ref{tab:benchmark}, TECSA-GNN achieves SOTA performance in predicting TE properties, a result attributable to its comprehensive encoding of crystal structures together with the incorporation of physicochemical statistical features. It is worth emphasising that the \textit{No Free Lunch} theorem applies equally to TECSA-GNN. We attempted to employ transfer learning~\cite{trans-learning} in order to train TECSA-GNN on the prediction of band gap $E_{\rm g}$ and formation energy $E_{\rm f}$. Interestingly, although these targets are in principle learnable, the resulting accuracy was not particularly impressive when compared with peer models. This phenomenon may be ascribed to the tendency of overparameterised models to overfit: for relatively simple tasks, an excessive number of parameters risks capturing spurious noise within the data, thereby impairing generalisation~\cite{mingard2025deep}. By contrast, TE transport properties are intrinsically complex physical quantities, intimately linked to crystal structure, electronic structure, and physicochemical attributes. The choice of architectural complexity should therefore be made with due consideration of the physical context. For energy-related quantities such as $E_{\rm g}$ and $E_{\rm f}$, ALIGNN~\cite{alignn} achieves SOTA performance, plausibly owing to its early incorporation of bond-angle information. Conversely, models relying solely on physicochemical statistical descriptors~\cite{LI2023106299} perform markedly worse. In other words, for energy prediction tasks, structural descriptors alone may suffice. This reflects the spirit of \textit{Occam's razor}~\cite{NEURIPS2024_2c572cad}: adding more information or employing more complex architectures is not necessarily beneficial.

A practical approach of assessing whether the model has effectively captured structural information is to employ $t$-SNE~\cite{maaten2008visualizing} to project the samples into a low-dimensional map. For this purpose, we extract the activations from the final fully connected MLP layer preceding the output node $\mathbf{\hat{y}}$ (see Fig.~\ref{fig:net}) of the pre-trained $S$ prediction model, and use them as sample representations. Here we chose the $S$ pre-trained model since its prediction accuracy was the highest and thus most effective for learning structural representations, while the sign of $S$ further indicates the doping type. A more detailed discussion of $\log(\sigma/\tau)$ and $\log(\kappa_{\rm e}/\tau)$ is provided in~\cite{supp}, and the $t$-SNE visualization of the $S$ model embeddings is shown in Fig.~\ref{fig:tsne-cs-s}.

From the perspective of crystal structures, the distribution of materials from different crystal systems in the $t$-SNE space exhibits three salient features: first, samples of the same system form continuous and adjoining clusters; second, structurally similar systems, such as \{Trigonal, Hexagonal\} or \{Cubic, Tetragonal, Orthorhombic\}, appear in close proximity; and third, the distributions of different doping types are arranged in an approximately symmetric manner. Continuity here refers to the clustering and adjacency of samples from the same crystal system, while proximity reflects the neighbourhood of structurally related systems. These phenomena indicate that the embeddings learned by TECSA-GNN contain sufficiently rich structural information, for only under this condition can samples with identical or similar structures appear close in the low-dimensional space~\cite{maaten2008visualizing,zhang2021out}. Symmetry, on the other hand, is manifested in the near left-right reflection of different systems. As shown in Fig.~\ref{fig:tsne-cs-s}(h), a possible explanation is that the horizontal axis encodes the qualitative separation of doping type, with $n$-type materials on the left and $p$-type on the right; intriguingly, some samples located in the upper region exhibit $S$ values opposite to their nominal doping type, a behaviour that may be associated with bipolar conduction phenomena~\cite{FOSTER201991,bipolar}.

We masked a set of special crystal structures in the dataset and evaluated the generalisation ability of TECSA-GNN on unseen configurations by fine-tuning the pretrained model. In practice, a total of 133 special structures (covering Full-Heusler (FH), Half-Heusler (HH), rocksalt (RS), and zinc blende (ZB) phases) were considered. Unlike in pretraining, these samples were evenly divided into training and testing subsets. Moreover, it was unnecessary to retrain all model parameters: the MLPs used for dimensional alignment and all GConv layers were kept frozen, and only the terminal regression head was fine-tuned. This strategy improves computational efficiency and preserves the stability of local feature extraction, while allowing the model to adapt specifically to these special structures without discarding prior knowledge.

Fig.~\ref{fig:fine-tune} compares the performance of pretrained and fine-tuned models at 300~K for four representative structures. Fine-tuning offers two principal advantages: first, it selectively improves predictions for these special structures, most notably by reducing outlier errors; second, it avoids the considerable computational cost of training from scratch, requiring only $\sim 50$ epochs in our case to achieve substantial accuracy gains. It should be noted, however, that the $R^2$ metric is inherently correlated with sample size~\cite{mckay1999sample}; consequently, the values reported in Fig.~\ref{fig:fine-tune} are lower than those in Fig.~\ref{fig:performance}. Across datasets of unequal size, MAE is a more dependable metric.

\subsection{Discovering potential thermoelectric materials}
\label{sec:htp}

Leveraging the pretrained TECSA-GNN model, we screened unlabeled data to identify promising candidates with potentially high PF values. The Materials Project~\cite{jain2013commentary} provides a vast repository of crystal structures along with DFT-computed band gaps. We applied the following criteria to filter candidate materials: (1) $0<E_{\rm g} <1.5$ [eV], (2) nonmagnetic, (3) free of transition metals, and (4) fewer than 20 atomic sites in the unit cell (${\rm n\_sites} < 20$). Criterion (1) is empirical in nature, as constraining the band gap $E_{\rm g}$ ensures that candidates lie within the semiconducting regime. Meanwhile, criteria (2), (3) and (4) are motivated primarily by the need to simplify subsequent DFT calculations. After excluding materials already present in our training dataset, we conducted DFT validations on three promising candidates at 300~K: [mp-1209957] NaTlSe$_2$, [mp-9897] Te$_3$As$_2$ and [mp-1207082] LiMgSb.

\begin{figure}[h]
    \centering
    \includegraphics[width=\columnwidth]{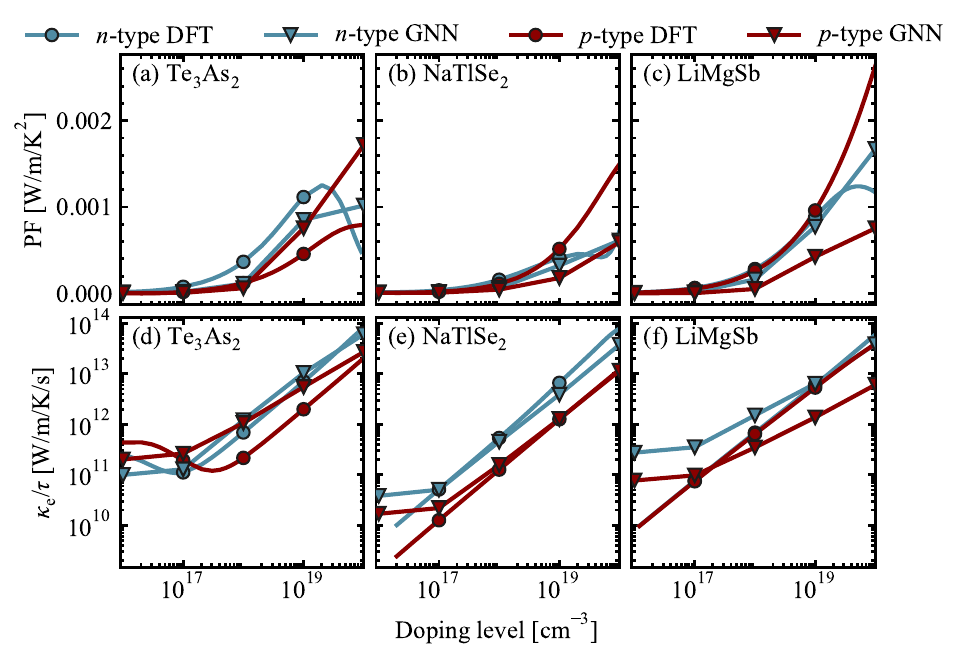}
    \caption{Comparison between TE transport properties predicted by the TECSA-GNN model and those obtained from DFT calculations. (a)-(c) display the PF values of Te$_3$As$_2$, NaTlSe$_2$, and LiMgSb, computed as $S^2\sigma/\tau\times10^{-14}$ rather than directly learned by TECSA-GNN. (d)-(f) show the corresponding $\kappa_{\rm e}/\tau$. Blue and red denote $n$-type and $p$-type carriers, respectively; inverted triangles represent the TECSA-GNN estimates, while circles represent the DFT-calculated values.}
    \label{fig:new-maters}
\end{figure}

Among them, NaTlSe$_2$ was previously reported by Ref.~\cite{Antunes2023}, albeit with analysis limited to its chemical formula, whereas our GNN-based approach enables structural evaluation at the crystal level. Recent studies suggest that a PF exceeding $6.8 \times 10^{-4}~{\rm [W/m/K^{2}]}$ (as in $p$-type SnSe)~\cite{su2023re} or $8 \times 10^{-4}~{\rm [W/m/K^{2}]}$ (as in $n$-type Sc$_{0.015}$Mg$_{3.185}$Sb$_{1.5}$Bi$_{0.47}$Te$_{0.03}$)~\cite{YU2024149051} can be considered qualitatively high at 300~K. In Fig.~\ref{fig:new-maters}, the $\kappa_{\rm e}$ of the candidates remains below 1~[W/m/K] across the considered doping range. From a TE perspective, however, this does not necessarily imply a low overall thermal conductivity, since $\kappa_{\rm ph}$ is typically the dominant contribution to heat transport~\cite{xu2019effect}. Nevertheless, when $\kappa_{\rm ph}$ itself is sufficiently low and becomes comparable in magnitude to $\kappa_{\rm e}$, the electronic contribution can no longer be neglected in evaluating the total thermal transport performance~\cite{wan2010development}.

Overall, TECSA-GNN shows good agreement with DFT calculations, particularly in the estimation of $S$. The predictions of $\sigma/\tau$ and $\kappa_{\rm e}/\tau$ are somewhat less accurate, yet as discussed in Sec.~\ref{sec:performance}, the dataset's approximation of the relaxation time $\tau$ inevitably introduces some uncertainty. Despite these errors, the efficiency of a pretrained AI model compared with \textit{ab initio} calculations makes it highly practical for high-throughput screening of new materials, enabling the rapid identification of promising candidates across vast, unexplored regions of compositional and structural space. In addition to NaTlSe$_2$, Te$_3$As$_2$, and LiMgSb, we also provide predictions for other unlabeled samples, as detailed in~\cite{supp}.

\subsection{Tracing thermoelectric behaviour from DFT insights}
\label{sec:dft-insights}

For the candidates identified through high-throughput screening, it is still premature to classify them directly as ``valuable materials'' solely on the basis of the DFT validation results. We expect that more rigorous evidence can be obtained through higher-fidelity theoretical calculations~\cite{YUE2024101517,doi:10.1021/jacs.5c15502,doi:10.1021/acsami.4c09043,doi:10.1021/acsami.4c02097}. Taking the candidates in Sec.~\ref{sec:htp} as examples, we performed DFT calculations of their band structures and electron localization functions (ELF)~\cite{savin1997elf} to trace the origin of their TE performance.

\begin{figure}[ht]
    \centering
    \includegraphics[width=\columnwidth]{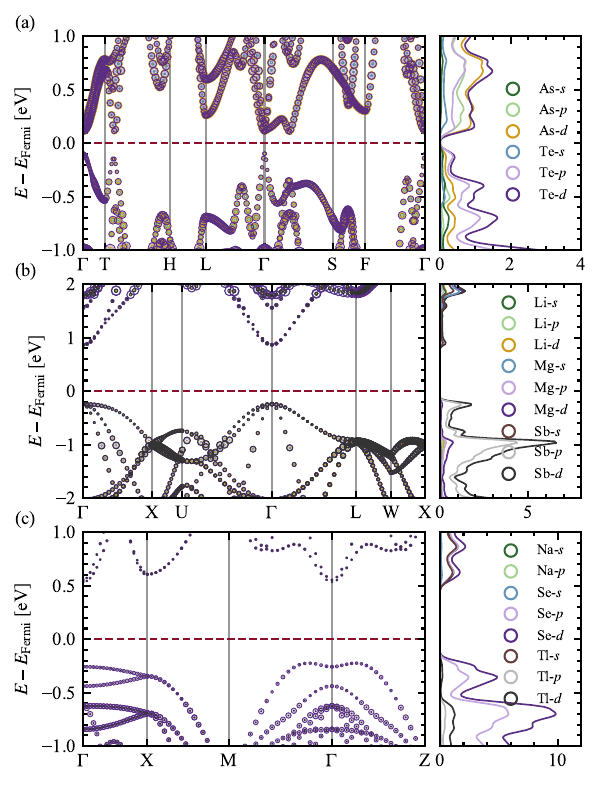}
    \caption{Projected band structure and PDOS. (a) Te$_3$As$_2$, (b) LiMgSb, (c) NaTlSe$_2$. In the left panel, each scatter represents the atomic-orbital projection weight at a given $k$ point and energy, with color denoting the atomic species and marker area proportional to the PDOS value of the corresponding atom and orbital, normalized by the maximum PDOS of that atom. Vertical lines indicate the high-symmetry $\mathbf{k}$-points, and the dashed horizontal line marks the Fermi level. The right panel shows the PDOS curves of each atom in the same colors as in the band plot.}
    \label{fig:band-struc}
\end{figure}

In Fig.~\ref{fig:band-struc}, the band structures of Te$_3$As$_2$, LiMgSb, and NaTlSe$_2$ exhibit distinct features. For Te$_3$As$_2$, both the conduction band minimum (CBM) and the valence band maximum (VBM) are located near $E_{\rm Fermi}$, indicating a small $E_{\rm g}$ and a narrow-gap character. The pronounced band dispersion around $E_{\rm Fermi}$ further implies a small $m^*$. According to Mott's formula for the Seebeck coefficient~\cite{PhysRevB.88.115128}, $S$ is closely related to the energy-dependent gradients of the density of states and scattering strength,
\begin{align}
    S = \frac{\pi^2 k_{\rm B}^2 T}{3e}
    \left.\frac{d\ln\sigma(E)}{dE}\right|_{E=E_{\rm Fermi}},
    \label{eq:mott-s}
\end{align}
where $\sigma(E) = e^2 N(E) v^2(E) \tau(E)$ represents the energy-dependent conductivity, with $\tau$ fixed at $10^{-14}$~s in our case. In Fig.~\ref{fig:band-struc}(a), the steeply dispersive bands of Te$_3$As$_2$ indicate slowly varying $v(E)$ and a relatively flat $N(E)$, resulting in a small conductivity gradient near $E_{\rm Fermi}$, i.e., $\frac{d\ln\sigma(E)}{dE} \propto \frac{d\left(\ln{N(E)v^2(E)\tau(E)}\right)}{dE}$, and consequently a low $S$~\cite{park2021optimal}. In fact, the excellent PF of Te$_3$As$_2$ arises from its exceptionally high $\sigma$, consistent with its narrow-gap nature. As illustrated in Fig.~\ref{fig:new-maters}(d)-(f), Te$_3$As$_2$ exhibits a markedly larger $\kappa_{\rm e}/\tau$ than LiMgSb and NaTlSe$_2$, which further corroborates this feature. 

According to Fig.~\ref{fig:band-struc}(b) and (c), LiMgSb and NaTlSe$_2$ exhibit several common features in their band structures. The VBM lies close to $E_{\rm Fermi}$, indicating hole-dominated conduction under intrinsic or lightly acceptor-doped conditions, consistent with the general trend that $p$-type TE materials usually achieve higher PF values~\cite{yang2015outstanding}. The cations contribute negligibly to the dominant states near the band edges; for instance, the projected density of states (PDOS) of Li in LiMgSb and Na in NaTlSe$_2$ is nearly zero, suggesting that these ions mainly act as electrostatic charge balancers rather than active contributors to transport. Unlike conventional thermoelectrics where cations participate in transport, Li/Na here play a unique charge-balancing role. The Na layers stabilize the Tl-Se framework and influence the geometry of Tl-Se bonds, thereby tuning the $d$-$p$ hybridization~\cite{PhysRevLett.109.037003} strength and the position of the VBM. This behavior aligns with previous findings that alkali cations in chalcogenide thermoelectrics act as electrostatic modulators rather than electronic contributors~\cite{exp-na}.  The valence bands of NaTlSe$_2$ are relatively flat, leading to large carrier effective masses and low mobility, and hence low $\sigma$, but the steep variation of the DOS near $E_{\rm Fermi}$ yields a large $\frac{d\ln\sigma(E)}{dE}$ and thus a high $S$. The flat-band nature and localization of states, however, suppress carrier mobility, limiting $\sigma$ and leading to a low electronic thermal conductivity $\kappa_{\rm e}$. Overall, NaTlSe$_2$ represents a system characterized by high $S$, low $\sigma$, and low $\kappa_{\rm e}$. Similarly, LiMgSb exhibits negligible Li orbital contributions near the valence and conduction bands, with Li ion acting mainly as an electrostatic stabilizer. The valence bands are dominated by Sb-$d$ orbitals, while the conduction bands are primarily Mg-$d$ in character, indicating that hole transport is governed by the Sb sublattice and electron transport occurs mainly within the Mg-Sb framework~\cite{gnanapoongothai2015first}. The strongly dispersive valence bands in LiMgSb indicate small $m^*$, which, according to the relation $\mu \propto 1/m^*$, lead to high carrier mobility and thus high $\sigma$. Conversely, flat valence bands (i.e., low dispersion) would correspond to large $m^*$ and reduced $\mu$~\cite{park2021optimal}. Meanwhile, the small DOS gradient results in a lower $S$, and the delocalized nature of the bands contributes to an elevated $\kappa_{\rm e}$. Overall, LiMgSb exhibits a balanced combination of transport properties, with moderate $S$, high $\sigma$, and intermediate $\kappa_{\rm e}$.

Apart from the electronic band structure, the ELF offers further theoretical insight into the distinct TE behaviours of the candidate materials. The ELF directly reflects the extent of electron localization or delocalization, a key factor governing carrier mobility and thus the electronic transport coefficients. Its values range from $\left[0,1\right]$: when ${\rm ELF}\approx 1$, it indicates strong electron localization, as in covalent bonds or lone-pair regions; when ${\rm ELF}\approx 0.5$, it corresponds to a delocalized, nearly homogeneous electron gas typical of metallic conduction; lower ELF values generally denote regions of sparse electron density, such as interionic voids.

\begin{figure*}[t]
    \centering
    \includegraphics[width=\textwidth]{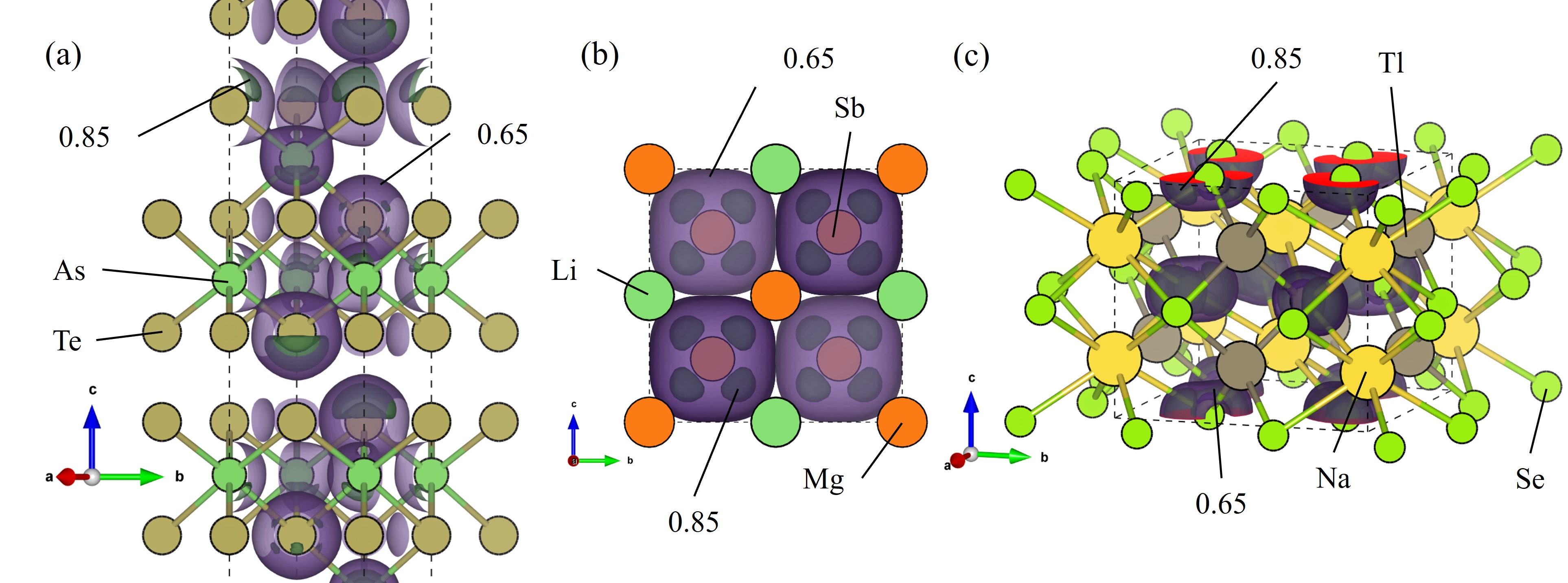}
    \caption{ELF spatial visualization. In (a) Te$_3$As$_2$, (b) LiMgSb, and (c) NaTlSe$_2$, the green and purple isosurfaces denote ELF value of 0.85 and 0.65, respectively.}
    \label{fig:edge-elf}
\end{figure*}

We calculated the ELF distributions of Te$_3$As$_2$, LiMgSb, and NaTlSe$_2$, as shown in Fig.~\ref{fig:edge-elf}. The atoms in Te$_3$As$_2$ exhibit a layered arrangement, and the ELF map in Fig.~\ref{fig:edge-elf}(a) further reveals a distinctive layered covalent framework. Specifically, a high degree of electron localisation is observed between Te and As atoms within the layers, indicating the formation of strongly directional covalent bonds. This is consistent with the PDOS in Fig.~\ref{fig:band-struc}(a), where both As and Te contribute markedly near the CBM and VBM. In contrast, Te-Te and As-As interactions show clear delocalisation, and almost no electron density is observed between the layers. The ELF characteristics of Te$_3$As$_2$ closely resemble those of SnSe, both displaying interlayer voids and enhanced metallicity~\cite{snse-layer}. Overall, this interplay of intralayer localisation and interlayer delocalisation gives rise to pronounced band dispersion (particularly along the $\Gamma$-L direction), leading to a reduced $m^*$, higher carrier mobility, and consequently an increased $\sigma$~\cite{williamson2017engineering}.

A distinctive feature of the 3D ELF of LiMgSb (see Fig.~\ref{fig:edge-elf}(b)) is the strong localization around Sb atoms, while Mg exhibits a delocalized character, with island-like localized regions distributed between Mg-Mg atoms. Li, on the other hand, is almost fully ionized. In general, when electrons are strongly confined to specific atoms or bonds in real space (such as in lone pairs or strong covalent bonds), their wavefunctions spread broadly in momentum space, resulting in weak band dispersion. This leads to a large effective mass $m^*$ and consequently reduced mobility~\cite{2015Multi}. In the case of LiMgSb, the Sb-$p$ electrons are localized but moderately hybridized with Mg, which manifests in the band structure as weakly dispersive, yet not completely flat, bands. This feature accounts for the relatively high $S$ and the lower $\sigma$ and $\kappa_{\rm e}$, respectively. Unlike LiMgSb, where island-like ELF regions appear between Mg-Mg atoms, Fig.~\ref{fig:edge-elf}(c) shows that the ELF of NaTlSe$_2$ is fully localized around Se atoms. In other words, NaTlSe$_2$ exhibits stronger localization and greater anion character than LiMgSb. This results in flat valence-band tops, as seen along the $\Gamma$-Z-P direction in Fig.~\ref{fig:band-struc}(c), which correlate with strong ELF localization ($>0.9$) around Se atoms. Such localization induces low band dispersion and hence large $m^*$, enhancing $S$ via $S \propto m^*$. Consequently, NaTlSe$_2$ exhibits the highest $S$ but the lowest $\sigma$ and $\kappa_{\rm e}$ among the three candidates. Taken together, these results reveal the intrinsic antagonism between $S$ and $\sigma$ in TE materials. Therefore, among the three compounds considered, LiMgSb, which maintains the most balanced combination of $S$ and $\sigma$, achieves the highest PF, as shown in Fig.~\ref{fig:new-maters}.

\subsection{Input sensitivity quantification and graph interpretation}

Interpretability~\cite{jiang2025interpretable} is a central theme across ``AI for Science''~\cite{miao2024toward}. In contrast to traditional AI applications such as computer vision, which are often dominated by black-box models and the sole pursuit of SOTA performance, ``AI for Science'' places greater emphasis on enabling AI models to reflect or even uncover underlying theoretical patterns. In practice, the complex topology and message-passing mechanisms make GNNs particularly challenging to interpret. Interpretation methods can be categorized into four groups~\cite{yuan2022explainability}: gradient-, perturbation-, decomposition-, and surrogate-based methods. In brief, gradient-based methods~\cite{pope2019explainability} compute the gradient of the output with respect to input features, where larger magnitudes indicate greater importance; perturbation-based methods~\cite{NEURIPS2020_e37b08dd} mask nodes, edges, or their combinations to observe changes in the output; decomposition-based methods~\cite{9547794} disentangle the prediction process to trace contributions of inputs across layers; and surrogate-based methods~\cite{NEURIPS2020_8fb134f2} approximate the GNN with a simpler but more interpretable model. 

In this work, we adopt different strategies to interpret the TECSA-GNN across multiple feature scales. Specifically, we use the {\sc GNNExplainer}~\cite{ying2019gnnexplainer} to assess the contribution of atomic embeddings $\mathbf{V}$ at the atomistic scale; for global crystal features $\mathbf{U}$, employ a lightweight MLP surrogate as a practical solution. Considering that among the TE quantities learned by TECSA-GNN only $S$ corresponds to an actual physical value, while both $\sigma/\tau$ and $\kappa_{\rm e}/\tau$ involve the relaxation-time approximation and thus contain inherent uncertainties, all subsequent interpretability analyses are conducted with $S$ as the research object.

\begin{figure}[t]
    \centering
    \includegraphics[width=\columnwidth]{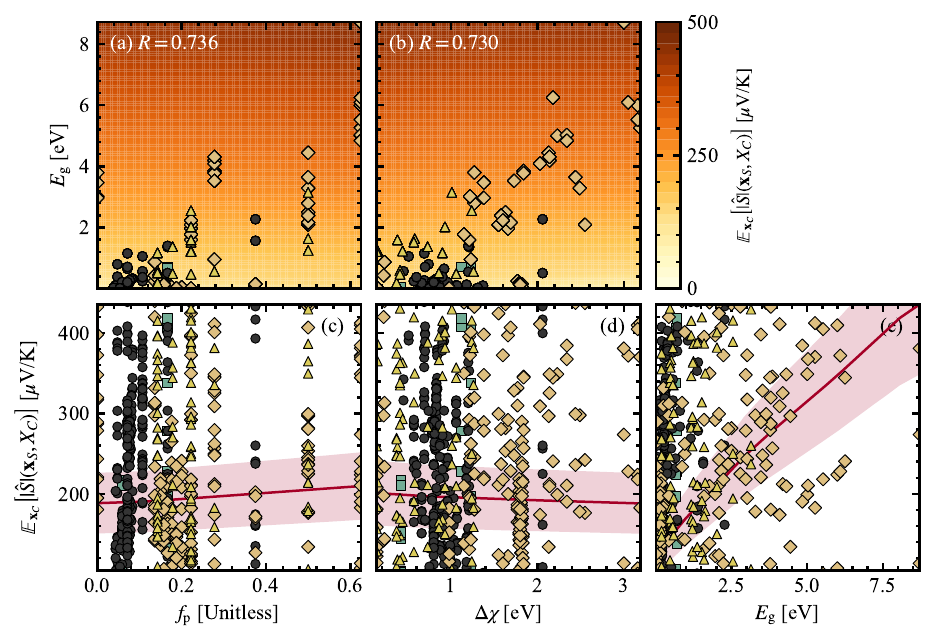}
    \caption{Partial dependence plots of the key global features for the estimated absolute Seebeck coefficient $| S \vert$. When the varying feature set $\mathbf{x}_S$ is fixed as (a) $\{E_{\rm g}, f_{\rm p}\}$, (b) $\{E_{\rm g}, \Delta\chi\}$, (c) $\{f_{\rm p}\}$, (d) $\{\Delta\chi\}$, and (e) $\{E_{\rm g}\}$, the estimated values of $\mathbb{E}_{\mathbf{X}_C}\left[|\hat{S}|\left(\mathbf{x}_S, \mathbf{X}_C\right)\right]$ are shown. In (a) and (b), $\mathbb{E}_{\mathbf{X}_C}\left[|\hat{S}|\left(\mathbf{x}_S, \mathbf{X}_C\right)\right]$ is a bivariate function obtained by uniform sampling within the ranges of each feature. In (c)-(e), $\mathbb{E}_{\mathbf{X}_C}\left[|\hat{S}|\left(\mathbf{x}_S, \mathbf{X}_C\right)\right]$ degenerates to a univariate function controlled solely by the variable on the $x$-axis. The red curve denotes the target dependence on the current variable, the pink band represents the $20\%$ uncertainty interval, and the marker shapes are consistent with Fig.~\ref{fig:fine-tune}, corresponding to HH, FH, ZB, and RS structures, with black circles representing other conventional structures.}
    \label{fig:pdp}
\end{figure}

\begin{figure*}[t]
    \centering
    \includegraphics[width=\textwidth]{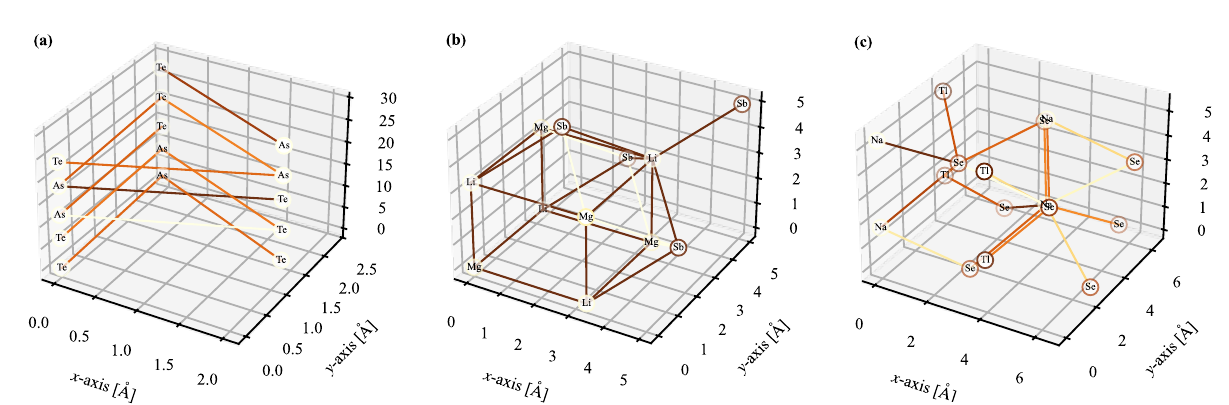}
    \caption{Atomic importance quantified by {\sc GNNExplainer} and the pairwise mean radial ELF values. In the unit cells of (a) Te$_3$As$_2$, (b) LiMgSb, and (c) NaTlSe$_2$, the front--back positions of the atoms are indicated by transparency, while the atomic importance is encoded by the colour gradient of the highlighted edges, with darker brown indicating a greater contribution to the TECSA-GNN decision. The diagrams also show the mean ELF values between pairs of atoms; only connections with an average ${\rm ELF} > 0.4$ are displayed, since values below this threshold suggest weak or negligible bonding.}
    \label{fig:node-impo}
\end{figure*}

Let us begin with a simple case. To investigate how the global features $\mathbf{U}$ influence the prediction of $S$, we constructed a $128 \times 128 \times 128$ MLP as a surrogate model, where crystal structure features were completely excluded. We first attempted to assess feature importance using \underline{SH}apley \underline{A}dditive \underline{E}xplanations (SHAP) method~\cite{lundberg2017unified}. However, apart from the band gap $E_{\rm g}$, which consistently exhibited the highest SHAP value, the intervention effects of other top-ranked features on the predicted $S$ were rather ambiguous (details can be found in the Supplemental Material~\cite{supp}). It is noteworthy that both $f_{\rm p}$, defined as the fraction of constituent elements possessing $p$-orbital valence electrons, and $\Delta \chi$, defined as the difference between the maximum and minimum electronegativities among the constituent elements, exhibit strong linear correlations with $E_{\rm g}$. Therefore, we retained $\left\{E_{\rm g}, f_{\rm p}, \Delta \chi\right\}$ for further discussion of their respective physical impacts on $S$. Fig.~\ref{fig:pdp} illustrates the partial dependence~\cite{friedman2001greedy} of $\left|\hat{S}\right|$ on $\{E_{\rm g}, f_{\rm p}, \Delta \chi\}$, where a subset of variables $\mathbf{x}_S$ is selected and the remaining features $\mathbf{X}_C$ are fixed at their mean values to remove irrelevant variation, such that the estimation of $\left|S\right|$ is written as $\mathbb{E}_{\mathbf{X}_C}\left[|\hat{S}|\left(\mathbf{x}_S, \mathbf{X}_C\right)\right]$. Here we focus on $\left|S\right|$ rather than $S$ itself in order to eliminate the sign change induced by the doping type, since in evaluating TE performance only the absolute magnitude of $S$, i.e., $\left|S\right|$, is of interest.

In Fig.~\ref{fig:pdp}, the most evident trend is that $E_{\rm g}$ shows a obvious positive correlation with $\left|S\right|$. The transport integral for $S$~\cite{russ2016organic,FRITZSCHE19711813} is defined as
\begin{align}
    S=-\frac{k_{\rm B}}{e}\int \left(\frac{E-E_{\rm Fermi}}{k_{\rm B}T}\frac{\sigma(E)}{\sigma}\right)\,dE,
    \label{eq:s-define}
\end{align}
where $k_{\rm B}$ is the Boltzmann constant, $e$ is the elementary charge, $E$ is the energy of the electronic state, $E_{\rm Fermi}$ is the Fermi level, $T$ is the absolute temperature, and the term $\sigma(E)/\sigma$ represents the normalized transport distribution function, i.e., the relative contribution of carriers at energy $E$ to the total electrical conductivity. From the perspective of Eq.~\eqref{eq:s-define}, $S$ may be simply interpreted as the weighted energy shift of carriers relative to $E_{\rm Fermi}$. When $E_{\rm g}$ is small, both valence- and conduction-band states near $E_{\rm Fermi}$ are thermally accessible, so that $\sigma(E)$ is distributed on both sides of $E_{\rm Fermi}$ and the integral term $(E-E_{\rm Fermi})$ in Eq.~\eqref{eq:s-define} cancels out, thereby reducing the average energy shift and lowering $\left|S\right|$. This corresponds to the so-called bipolar conduction phenomenon~\cite{gong2016investigation}. A larger $E_{\rm g}$ benefits $S$ by mitigating thermal excitation, suppressing the minority-carrier population, and thus weakening bipolar conduction~\cite{PhysRevB.111.045203}.

As illustrated in Fig.~\ref{fig:pdp}(c) and (d), the effects of $f_{\rm p}$ and $\Delta \chi$ on $\left|\hat{S}\right|$ are less direct than that of $E_{\rm g}$, and such features typically influence model predictions only through interactions with others~\cite{molnar2025}. Nevertheless, Fig.~\ref{fig:pdp}(a) and (b) reveal a rough yet interesting trend: larger $f_{\rm p}$ and $\Delta \chi$ tend to be associated with higher $E_{\rm g}$, thereby indirectly enhancing $\left|\hat{S}\right|$. Electronegativity $\chi$ is a measure of an element's ability to attract electrons within a compound, and its contribution to $E_{\rm g}$ can be divided into covalent and ionic parts: the former is governed by the average electronegativity $\bar{\chi}$ of the compound, while the latter is determined by $\Delta \chi$~\cite{chi-gap}. Specifically, D.~Adams~\cite{adams1974introduction} proposed the following definition of $E_{\rm g}$,
\begin{align}
    E^2_{\rm g}=E^2_{\rm h}+C^2,
\end{align}
where $E^2_{\rm h}$ denotes the covalent contribution to the $E_{\rm g}$, determined by $\bar{\chi}$, and $C$ is the charge-transfer energy, directly correlated with $\Delta \chi$. 

By contrast, the role of $f_{\rm p}$ in determining $E_{\rm g}$ is less clear. Ref.~\cite{fp-bandgap} speculated that increasing $f_{\rm p}$ enhances the interaction at the band edges and thereby enlarges $E_{\rm g}$, though no conclusive evidence was provided. Inspired by Ref.~\cite{chi-gap}, we conjecture that $f_{\rm p}$ influences $E_{\rm g}$ indirectly via its effect on $\chi$. In Fig.~\ref{fig:pdp}(b), within the approximate range $\Delta \chi<1.4$~[eV], $E_{\rm g}$ and $\Delta \chi$ are in fact roughly anticorrelated; compounds in this regime tend to be covalent~\cite{sproul2001electronegativity} (strictly speaking $\Delta \chi<1.7$~[eV]), whereas larger $\Delta \chi$ favours ionic bonding. In covalent compounds, $p$-orbital electrons dominate the bonding and antibonding states, and the band gap arises from the energy difference between the valence-band maximum (typically $p$-like) and the conduction-band minimum (of $s/p$ or $s/d$ hybrid character). When $f_{\rm p}$ is high, the valence-band maximum is pushed upward by $p$ states, often reducing $E_{\rm g}$~\cite{choi2009tunability}. In ionic compounds, however, a large $\Delta \chi$ enhances the contribution of anion $p$ states to the valence band, which raises $f_{\rm p}$ and leads to a larger gap~\cite{MAKINO199455}. Thus, $\Delta \chi$ may be the more direct driver of band-gap enlargement, while an increase in $f_{\rm p}$ may merely reflect the concomitant effect of increasing $\Delta \chi$.

The sensitivity analysis with respect to $\mathbf{U}$ applies to the entire materials space, whereas for specific candidates, e.g., Te$_3$As$_2$, LiMgSb, and NaTlSe$_2$, one may quantify the influence of individual atoms on the model's decision at the atomic scale, denoted by $\mathbf{V}$. This can be achieved through single-node explanations provided by {\sc GNNExplainer}~\cite{ying2019gnnexplainer}. The detailed implementation of {\sc GNNExplainer} will be presented in Sec.~\ref{sec:gnnexp}. Briefly, it evaluates the importance of a node by comparing the change in the model's output before and after masking that node. Using this approach, we performed atomic-node importance analyses for Te$_3$As$_2$, LiMgSb, and NaTlSe$_2$, and found that atomic importance is, to some extent, correlated with the degree of electronic localisation and the contribution of each element to the PDOS. In Fig.~\ref{fig:node-impo}(a)-(c), the normalized importance values assigned by {\sc GNNExplainer} to each atom in the prediction of $S$ are mapped by color. 

Fig.~\ref{fig:node-impo}(a) shows that all atoms in Te$_3$As$_2$ exhibit zero importance, indicating that the atomic features $\mathbf{V}$ of Te$_3$As$_2$ play an almost negligible role in the model's estimation of $S$ for this compound. 
From a physical standpoint, this phenomenon may stem from the delocalisation of electronic states, leading the decisive variables to become global in nature. Although the electrons around Te and As atoms are highly localised (high ELF), orbital hybridisation occurs between layers, resulting in wavefunction delocalisation along the interlayer direction in band space. According to Eq.~\eqref{eq:s-define}, the Seebeck coefficient $S$ depends on the asymmetry of the bands near the Fermi energy $E_{\rm Fermi}$ and on the global transport function,
\begin{align}
    \sigma(E) \sim \sum_{\mathbf{k},n} v^2_{n\mathbf{k}}\, \tau_{n\mathbf{k}}\, \delta(E - E_{n\mathbf{k}}),
\end{align}
where $n$ denotes the band index, $\mathbf{k}$ is the wavevector in the Brillouin zone, $v_{n\mathbf{k}}$ is the electron group velocity, $\tau_{n\mathbf{k}}$ is the relaxation time, and $E_{n\mathbf{k}}$ is the band energy. This expression shows that the energy-dependent conductivity $\sigma(E)$ represents a weighted accumulation of transport contributions from all bands and $\mathbf{k}$ points, governed jointly by the squared electron velocity, scattering lifetime, and density of states. In many semimetals or narrow-gap systems where the Fermi surface is dominated by highly delocalised Bloch states, $\sigma(E)$ is determined by the overall band structure and the average velocity or relaxation time. Local atomic perturbations are often ``diluted'' by such averaging effects, resulting in only minor changes to the overall shape of $\sigma(E)$~\cite{PhysRevB.87.195205}. Nevertheless, when local perturbations introduce resonant or trap states or significantly modify the local transition matrix elements, thereby changing the density of states or carrier lifetimes near the $E_{\rm Fermi}$, single-site variations can still exert a pronounced influence on $\sigma(E)$ and on derived physical quantities such as $S$~\cite{PhysRevB.87.195205}.

In the semiconductors MgLiSb and NaTlSe$_2$, the relative importance of individual atoms appears to be closely linked to their specific roles in the electronic band structure. Qualitatively, as shown in Fig.~\ref{fig:node-impo}(b), the Sb atoms in MgLiSb exhibit higher importance than Li and Mg; whereas in Fig.~\ref{fig:node-impo}(c), Se shows slightly greater importance than Tl, with Na consistently being the least important. Examining the band structures in Fig.~\ref{fig:band-struc}(b) and (c), one finds that these importance values qualitatively reflect the dominant orbital contributions of different elements: in LiMgSb, the VBM is almost entirely governed by Sb, while Li and Mg show negligible features in their PDOS near either the CBM or the VBM. In contrast, in NaTlSe$_2$, Se and Tl jointly determine the shape of the CBM, whereas the VBM is almost exclusively controlled by Se.

\section{Discussion}

The proposed multiscale GNN model, TECSA-GNN, accurately captures the complex and nonlinear relationships between crystal structures and TE transport properties. By integrating global compositional features with atomic, bond, and angular representations within a unified message-passing framework, the model achieves outstanding predictive performance on a large DFT-computed dataset. Among the three key TE descriptors, the Seebeck coefficient $S$ shows the most reliable prediction, which arises from the model's ability to effectively learn intrinsic connections between electronic structure and band topology.

Beyond improved accuracy, TECSA-GNN reveals physically consistent trends that align with conventional transport theory. The model identifies a positive correlation between $|S|$ and the band gap $E_g$, reflecting the suppression of bipolar conduction in wide-gap materials, consistent with the Mott relation. Interpretability analyses further indicate that for systems dominated by localized states such as NaTlSe$_2$ and LiMgSb, the network assigns greater importance to atoms associated with lone-pair orbitals or covalent distortions, in agreement with the PDOS and ELF results. In contrast, for delocalized or narrow-gap systems such as Te$_3$As$_2$, the importance distribution becomes more uniform, suggesting that the model's decisions rely on overall band dispersion rather than localized chemical motifs. These consistent patterns demonstrate that the learned representations encode not only numerical correlations but also the underlying physical mechanisms governing TE transport.

The multiscale architecture of TECSA-GNN plays a decisive role in both performance and interpretability. Hierarchical message propagation from angular to bond to atomic levels allows the model to aggregate short-range geometrical distortions and long-range chemical contexts, which are essential for transport properties determined by orbital hybridization and band connectivity. The inclusion of composition-level descriptors further enhances the transferability of the model, enabling efficient fine-tuning across different chemical families. This multilevel coupling effectively bridges the gap between empirical descriptors and first-principles calculations, establishing a generalizable paradigm for high-throughput screening.

However, some intrinsic limitations of our work should be emphasised. The assumption of a constant relaxation time $\tau=10^{-14}$ s, while common in transport databases, cannot reflect variations among materials caused by defects, temperature, or phonon scattering, which limits the absolute prediction of $\sigma$ and $\kappa_{\mathrm e}$. Furthermore, reducing anisotropic tensors to scalar averages may overlook directional effects in layered or low-symmetry systems. Although multiple interpretability approaches such as {\sc GNNExplainer}, perturbation analysis, and surrogate MLP models highlight meaningful features, the distinction between statistical relevance and true causality should be treated cautiously since nonlinear feature entanglement may obscure direct physical attribution.

Future improvements can proceed along four directions. First, incorporating relaxation-time models based on phonon spectra or empirical scattering approximations would enhance the physical realism of the predictions. Second, applying uncertainty quantification through ensemble or Bayesian techniques~\cite{rahaman2021uncertainty} could provide confidence intervals for large-scale material screening. Third, employing equivariant GNNs~\cite{zhong2023transferable} that preserve tensorial symmetry would allow direct prediction of anisotropic transport tensors. Lastly, coupling these developments with active learning and iterative DFT validation would establish an efficient closed-loop framework for material discovery~\cite{sheng2020active}.

In summary, TECSA-GNN successfully establishes a physically consistent mapping between atomic geometry and TE transport by encoding multiscale structural and electronic information. Its strong predictive power and interpretability provide a solid foundation for integrating machine learning with first-principles calculations, offering a promising pathway toward accelerated discovery of high-performance TE materials and deeper understanding of structure–property relationships.

\section{Methods}

\subsection{Creating the dataset}
\label{sec:dataset}
The training dataset employed in this work originates from the DFT calculations of Ref.~\cite{ricci2017ab}. The original dataset comprises a total of 47,737 materials, each evaluated under both $n$- and $p$-type doping conditions, 5 carrier concentrations ranging from $10^{16}$ to $10^{20}$ cm$^{-3}$, and 13 temperature points spanning 100~K to 1300~K. For each setting, the TE properties $S$, $\sigma$, and $\kappa_{\rm e}$ are reported. After removing samples with missing information, duplicates, elemental single-component systems, and those for which feature construction failed, a total of 16,637 valid entries were retained. It is worth noting that, owing to crystalline anisotropy, the TE properties are provided in the form of $3 \times 3$ matrices, 
\begin{align}
    \bf{y} = \left( {\begin{array}{*{20}{c}}
{{y_{xx}}}&{{y_{xy}}}&{{y_{xz}}}\\
{{y_{yx}}}&{{y_{yy}}}&{{y_{yz}}}\\
{{y_{zx}}}&{{y_{zy}}}&{{y_{zz}}}
\end{array}} \right).
\end{align}

For the sake of simplification, we consider the normalized trace of $\mathbf{y}$, defined as $y=\tfrac{1}{3}{\rm Tr}[\mathbf{y}]$, which yields a scalar quantity that is subsequently adopted as the learning target of the model. 

An important subtlety is that the conductivity values reported in the dataset do not, in fact, correspond to the absolute conductivity but rather to the conductivity normalised by the relaxation time, namely the intrinsic conductivity $\sigma/\tau$, where $\tau=10^{-14}~{\rm s}$. This quantity is determined solely by the electronic structure of the material, e.g., the carrier concentration $n$ and the effective mass $m^*$~\cite{kitamura2015derivation}. The relaxation time $\tau$, meanwhile, encapsulates the underlying scattering processes in the crystal, governed by a multitude of factors including temperature, defects, and impurities~\cite{smith1982frequency,ahmad2010energy}, and its explicit computation remains notoriously difficult. The adoption of $\sigma/\tau$ is therefore a widely employed simplification: whilst intrinsic conductivity is not strictly identical to the measurable conductivity, it nonetheless serves as a meaningful proxy for assessing the propensity of a material to behave as either a good conductor or an insulator~\cite{hu2023intrinsic}.

\subsection{Transforming the crystal structures into graph representations}
\label{sec:trans-feat}

In Eq.~\eqref{eq:graph-define}, each crystal is represented in terms of its topological and physicochemical characteristics at four hierarchical levels: global, atomic, bond, and angular. The global feature vector $\mathbf{U}$ is constructed using the Magpie descriptors~\cite{ward2016general}, which provide elemental physicochemical properties such as melting point and electronegativity based on the constituent elements, and compute statistical quantities such as the mean, variance, and extrema, weighted by the stoichiometric ratios. Similarly, the atomic feature vector $\mathbf{V}$ is derived following the same scheme; however, for individual atoms, statistical measures are redundant. Therefore, $\mathbf{V}$ records the intrinsic physicochemical properties of each element, whereas $\mathbf{U}$ captures the statistical aggregates. In practice, the normalized atomic coordinates are concatenated into the atomic feature vector, $\mathbf{V}_i = \mathbf{r}_i^{\mathrm{norm}} \oplus f_{\mathrm{elem}}(a_i)$.

For each atomic pair $\left(a_i,a_j\right)$, the scalar distance is computed from their coordinates $\left\{\mathbf{r}_i,\mathbf{r}_j\right\}$ as $d_{ij}=\|\mathbf{r}_i-\mathbf{r}_j\|$. In addition, the normalized bond direction vector is defined as $\widehat{\mathbf{u}}_{ij} = \frac{\mathbf{r}_j - \mathbf{r}_i}{\| \mathbf{r}_j - \mathbf{r}_i \|}$, which is concatenated with the RBF embedding of bond length to form the final bond feature.

For each atomic triplet $\left(a_j,a_i,a_k\right)$, we define the directional vectors $\left\{\mathbf{u}=\mathbf{r}_j-\mathbf{r}_i,\mathbf{v}=\mathbf{r}_k-\mathbf{r}_i\right\}$, and the bond angle is given by $\theta_{jik}=\arccos{\left(\frac{\mathbf{u} \cdot \mathbf{v}}{\|\mathbf{u}\|\|\mathbf{v}\|}\right)}$. Both $\left\{d_{ij},\theta_{jik}\right\}$ are further transformed into vectors by the RBF expansion as
\begin{align}
\begin{cases}
    \mathbf{V}_{ij}=\exp\left(-\frac{(d_{ij}-\mu_k)^2}{\sigma_d^2}\right),\\
    \mathbf{\Theta}_{jik}=\exp\left(-\frac{(\theta_{jik}-\nu_m)^2}{\sigma_{\theta}^2}\right).
\end{cases}
\end{align}

In our implementation, the RBF expansion for bond lengths covers the range $\left[0,8\right]$~\AA, with center parameters $\mu_k=\mathrm{linspace}(0,8,8)$, a center spacing of $\Delta_d=8/7\approx1.14$~\AA, and a width parameter of $\sigma_d=1.5\Delta_d\approx1.71$~\AA. For bond angles, the RBF expansion spans $\left[0,\pi\right]$, with centers $\nu_m=\mathrm{linspace}(0,\pi,8)$, spacing $\Delta_\theta=\pi/7\approx0.45$, and width $\sigma_\theta=1.5\Delta_\theta\approx0.67$. This Gaussian-type expansion constructs a set of smooth and differentiable basis functions in continuous geometric space, effectively enabling the neural network to analytically capture structural variations within local geometric neighborhoods (receptive fields). It finely resolves local differences in bond lengths and angles, while the continuous kernel representation enhances smoothness and transferability in learning nonlinear relationships~\cite{NIPS2017_303ed4c6,Gasteiger2020Directional}.

\begin{algorithm}[h]
\caption{Representing Crystal Graph}
\label{alg:graph_construction}
\SetAlgoLined
\KwIn{Crystal structure $\mathcal{C}$, global descriptors $\mathbf{U}$, band gap $E_{\rm g}$, doping type $t\in\{0,1\}$ ($t=0$ for $n$-type), doping level $n\in\{10^{16},10^{17},\dots,10^{20}\}$ cm$^{-3}$}
\KwOut{Multiscale graph $\mathcal{G}=(\mathbf{U},\mathbf{V},\mathbf{E},\mathbf{\Theta})$}

$\mathcal{A}\gets\{a_i\}_{i=1}^{N}$ extracted from $\mathcal{C}$\;

\For{$i\gets1$ \KwTo $N$}{
  $\mathbf{V}_i\gets \mathbf{r}_i^{\mathrm{norm}} \oplus f_{\mathrm{elem}}(a_i)$\;
  $\mathcal{N}_k(i)\gets k\text{-NN}(a_i;\mathcal{C})$ with periodic boundary\;
}

\For{$i\gets1$ \KwTo $N$}{
  \ForEach{$j\in\mathcal{N}_k(i)$}{
    $d_{ij}\gets \mathrm{dist}(a_i,a_j)$\;
    $\widehat{\mathbf{u}}_{ij}\gets\dfrac{\mathbf{r}_j-\mathbf{r}_i}{\|\mathbf{r}_j-\mathbf{r}_i\|}$\;
    $\mathbf{E}_{ij}\gets \widehat{\mathbf{u}}_{ij}\oplus\phi_{\mathrm{RBF}}(d_{ij})$\;
  }
}

\For{$i\gets1$ \KwTo $N$}{
  \ForEach{$(j,k)\in\mathcal{N}_k(i)^2,\ j\neq k$}{
    $\theta_{jik}\gets \angle(a_j,a_i,a_k)$\;
    $\mathbf{\Theta}_{jik}\gets \phi_{\mathrm{RBF}}(\theta_{jik})$\;
  }
}

$\mathbf{U}\gets f_{\mathrm{global}}(\mathcal{C})\oplus n\oplus t\oplus E_{\rm g}(\mathcal{C})$\;

$\mathbf{V}\gets\{\mathbf{V}_i\}_{i=1}^{N}$\;
$\mathbf{E}\gets\{\mathbf{E}_{ij}\}_{i\in[N],\,j\in\mathcal{N}_k(i)}$\;
$\mathbf{\Theta}\gets\{\mathbf{\Theta}_{jik}\}_{i\in[N],\,j,k\in\mathcal{N}_k(i),\,j\neq k}$\;

$\mathcal{G}\gets(\mathbf{U},\mathbf{V},\mathbf{E},\mathbf{\Theta})$\;

\Return{$\mathcal{G}$}
\end{algorithm}
Alg.~\ref{alg:graph_construction} illustrates how TECSA-GNN constructs graph representations. This process compresses the complex crystal structure into a machine-readable embedding vector.

\subsection{Evaluating the atom/node importance via {\sc GNNExplainer}}
\label{sec:gnnexp}
For a pretrained GNN model $f_{\theta}(\mathcal{G}) \to y$, {\sc GNNExplainer}~\cite{ying2019gnnexplainer} aims to identify a minimal subset of atomic nodes whose presence is sufficient to preserve the original prediction, i.e., $f_{\theta}(\mathcal{G}_{\rm s}) \approx f_{\theta}(\mathcal{G})$. Since direct optimisation over discrete node selections is NP-hard, {\sc GNNExplainer} introduces a continuous node-masking mechanism~\cite{mishra2020node}. Each atom $a_i$ is associated with a learnable mask parameter $M_i \in [0,1]$ that controls its contribution to message passing. The masked graph is represented as
\begin{align}
\begin{cases}
\mathbf{H}' = \mathbf{H} \odot \sigma(\mathbf{M}_v), \\
\hat{y}_t = f_{\theta}(\mathcal{G},\, \mathbf{H}'),
\end{cases}
\end{align}
where $\mathbf{H} \in \mathbb{R}^{N\times d}$ is the node feature matrix with $N$ atoms and $d$-dimensional descriptors, $\mathbf{M}_v \in \mathbb{R}^{N}$ is a learnable continuous node mask, $\sigma(\cdot)$ is the sigmoid function constraining each mask value to $[0,1]$, and $\hat{y}_t$ is the predicted output for the masked graph. To enforce predictive fidelity and encourage sparsity and smoothness, the following loss is defined:
\begin{align}
\begin{cases}
\mathcal{L}_{\mathrm{pred}} &= 
\left[f_{\theta}(\mathcal{G}, \mathbf{H}')- f_{\theta}(\mathcal{G}, \mathbf{H})\right]^2, \\
\mathcal{L}_{\mathrm{reg}} &= 
\alpha \|\mathbf{M}_v\|_1 + \beta \Bigl[-\sigma(\mathbf{M}_v)\log\left(\sigma(\mathbf{M}_v)\right)\\& - \left(1-\sigma(\mathbf{M}_v)\right)\log\left(1-\sigma(\mathbf{M}_v)\right)\Bigr], \\
\min\limits_{\mathbf{M}_v} \mathcal{L} &= \mathcal{L}_{\mathrm{pred}} + \mathcal{L}_{\mathrm{reg}},
\end{cases}
\end{align}
where $\alpha$ and $\beta$ control the sparsity and smoothness of the mask, respectively. The mask parameters are updated via gradient descent:
\begin{align}
\mathbf{M}_v \leftarrow \mathbf{M}_v - \eta \frac{\partial \mathcal{L}}{\partial \mathbf{M}_v}.
\end{align}

After convergence, the learned mask becomes sparse, isolating the critical atoms that dominate the model's prediction. 
Nodes with higher mask values indicate the regions of chemical or structural significance, forming the local explanatory subgraph $\mathcal{G}_{\rm s}$.

\subsection{DFT calculations}

DFT calculations in this work were performed using the Vienna \textit{Ab-initio} Simulation Package (VASP)~\cite{PhysRevB.54.11169}. The projector augmented wave (PAW) method was employed to describe the electron-ion interactions~\cite{PhysRevB.54.11169,PhysRevB.59.1758}. The exchange-correlation functional was treated within the generalized gradient approximation (GGA) as parameterized by Perdew-Burke-Ernzerhof (PBE)~\cite{PhysRevLett.77.3865}. A plane-wave cutoff energy of 500 eV was used, and the electronic self-consistent iteration was converged to within $10^{-8}$ eV. Structural optimizations were carried out using $\mathbf{k}$-point meshes generated with \texttt{VASPKIT}~\cite{wang2021vaspkit}, with a $\mathbf{k}$-spacing of $0.03$ \AA$^{-1}$. TE transport coefficients were evaluated by solving the linearized Boltzmann transport equation as implemented in the \texttt{BoltzTraP2} code~\cite{madsen2018boltztrap2}. 

It should be clarified that in Fig.~\ref{fig:band-struc}, the DFT-calculated $E_{\rm Fermi}$ may be subject to some deviation, and shifting $E_{\rm Fermi}$ to the VBM provides a more meaningful reference~\cite{PhysRevB.95.165204}; in addition, compared with PBE, the Heyd--Scuseria--Ernzerhof (HSE) functional~\cite{hse} generally provides higher accuracy (at greater computational cost). More detailed discussion is provided in the Supplemental Material~\cite{supp}.

\begin{acknowledgments}
We would like to acknowledge the National Key Research and Development Program of China (Grant No.~2023YFB4603800), the financial support from the project of the National Natural Science Foundation of China (Grants No.~12404062, No.~12474093, No.~12302220, No.~12374091), Translational Medicine and Interdisciplinary Research Joint Fund of Zhongnan Hospital of Wuhan University (Grant No.~ZNJC202235), and the Fundamental Research Funds for the Central Universities (Grant No.~2042023kf0109). The numerical calculations in this work have been done on the supercomputing system in the Supercomputing Center of Wuhan University.
\end{acknowledgments}

\section*{Data Availability}
The data that support the findings of this article are publicly available~\cite{Zeng_TECSA-GNN}.





\bibliographystyle{apsrev4-2}
\bibliography{ref}

\clearpage

\onecolumngrid
\begin{center}
\subsection*{\large \textbf{Supplemental Material for} \\\textit{Learning Thermoelectric Transport from Crystal Structures via Multiscale Graph Neural Network}}
\end{center}

\setcounter{table}{0}
\setcounter{figure}{0}
\setcounter{equation}{0}
\setcounter{subsection}{0}
\renewcommand{\thetable}{S\arabic{table}}
\renewcommand{\thefigure}{S\arabic{figure}}
\renewcommand{\theequation}{Supplementary Eq.~\arabic{equation}}
\renewcommand{\thesubsection}{Supplementary Note \arabic{subsection}}

\subsection{Quantifying the effect of $E_{\rm g}$ calculation methods on electronic transport prediction}
\label{sup:1}
In the TECSA-GNN workflow, the band gap $E_{\rm g}$ is the only input feature required as prior knowledge. For the training dataset, $E_{\rm g}$ is obtained from generalized gradient approximation (GGA)~\cite{PhysRevLett.77.3865} calculations. However, computing $E_{\rm g}$ for each candidate during high-throughput screening is impractical. Therefore, we consider employing a pre-trained model to assist in estimating $E_{\rm g}$.
\begin{figure}[h]
    \centering
    \includegraphics[width=0.35\textwidth]{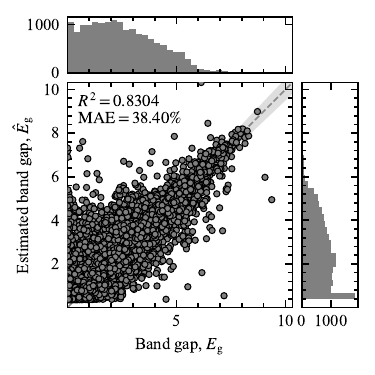}
    \caption{Parity plot of the predicted $E_{\rm g}$. Here, the $x$-axis denotes the DFT-calculated $E_{\rm g}$, while the $y$-axis represents the CGCNN-predicted values. The diagonal line serves as the reference, with grey shaded regions indicating the 95\% confidence intervals. The histograms along the top and right display the distribution of samples within each range. $\hat{E}_{\rm g}$ is the estimated band gap.}
    \label{fig:cgcnn-eg}
\end{figure}

The crystal graph convolutional neural network (CGCNN)~\cite{cgcnn} is a baseline GNN well suited for learning fundamental material properties. In our work, we employ it to predict the band gap $E_{\rm g}$ of candidate materials, followed by TECSA-GNN to estimate their electronic transport coefficients. As shown in Fig.~\ref{fig:cgcnn-eg}, the prediction error exhibits a clear distributional dependence, primarily arising from samples with low $E_{\rm g}$. Considering that materials with high PF values are typically semiconductors with relatively low $E_{\rm g}$, we regard $\hat{E}_{\rm g}$ as a compromise between efficiency and accuracy for high-throughput screening, rather than a full replacement of DFT in determining the final transport properties of materials.

\begin{figure}[h]
    \centering
    \includegraphics[width=0.85\textwidth]{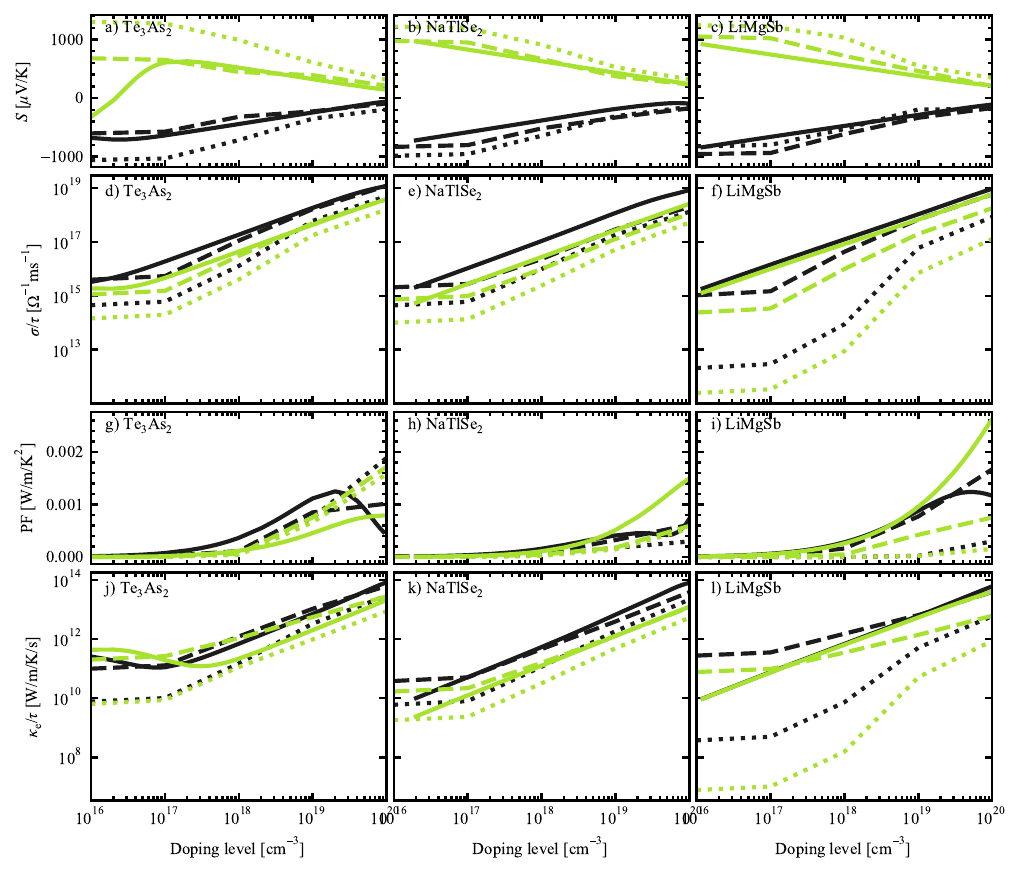}
    \caption{Doping-concentration-dependent electronic transport coefficients of the candidate materials at 300~K. Here, black denotes $n$-type and green denotes $p$-type. Each column corresponds to one compound. Solid, dashed, and dotted lines represent the DFT-calculated values, TECSA-GNN with DFT-calculated $E_{\rm g}$, and TECSA-GNN with CGCNN-predicted $E_{\rm g}$, respectively.}
    \label{fig:cgcnn-dft-eg-pred}
\end{figure}

To further quantify the impact of using $\hat{E}_{\rm g}$ on the predictions, Fig.~\ref{fig:cgcnn-dft-eg-pred} compares the transport property curves obtained from TECSA-GNN when supplied with either $E_{\rm g}$ or $\hat{E}_{\rm g}$ as input, alongside the results from DFT \textit{ab initio} calculations.

As shown in Fig.~\ref{fig:cgcnn-dft-eg-pred}, the absence of DFT-calculated band gaps as input leads to varying degrees of deviation in the model predictions. For $\hat{S}_{\hat{E}_{\rm g}}$, the discrepancy from the true values remains acceptable, and all three prediction curves converge well as the carrier concentration increases. However, for $\hat{\sigma}_{\hat{E}_{\rm g}}$ and $\hat{\kappa}_{{\rm e}\text{-}\hat{E}_{\rm g}}$, the errors at low carrier concentrations reach two to three orders of magnitude, and even at higher concentrations the deviations remain at one to two orders of magnitude. To ensure the reliability of TECSA-GNN predictions, the optimal strategy is to provide DFT-calculated band gaps, followed by the use of a pre-trained band-gap model with higher predictive accuracy.

\subsection{Hyperparameter settings for TECSA-GNN training}
\label{sup:2}
The hyperparameters used to train TECSA-GNN are listed in Tab.~\ref{tab:tecsa-hyperparams}.

\begin{table}[h]
  \centering
  \caption{Hyperparameter settings used for TECSA-GNN training for each target.}
  \label{tab:tecsa-hyperparams}
  \small
  \begin{tabular}{llll}
    \toprule
    \textbf{Hyperparameter} & $S$ & $\log(\sigma)$ & $\log(\kappa_{\rm e})$ \\
    \midrule
    Batch size & 80 & 128 & 128 \\
    Initial learning rate & $1\times10^{-2}$ & $1\times10^{-4}$ & $1\times10^{-4}$ \\
    Training epochs & 200 & 2500 & 2500 \\
    Hidden dimension & 256 & 256 & 256 \\
    Number of GConv layers & 5 & 5 & 5 \\
    Optimizer & Adam~\cite{zhang2018improved} & Adam~\cite{zhang2018improved} & Adam~\cite{zhang2018improved} \\
    Weight decay & $1\times10^{-5}$ & $1\times10^{-5}$ & $1\times10^{-5}$ \\
    Loss function & MSE & MSE & MSE \\
    \botrule
  \end{tabular}
\end{table}

\subsection{Visualizing model embeddings of $\log(\sigma/\tau)$ and $\log(\kappa_{\rm e}/\tau)$}
\label{sup:3}
\begin{figure}[h]
    \centering
    \includegraphics[width=0.75\textwidth]{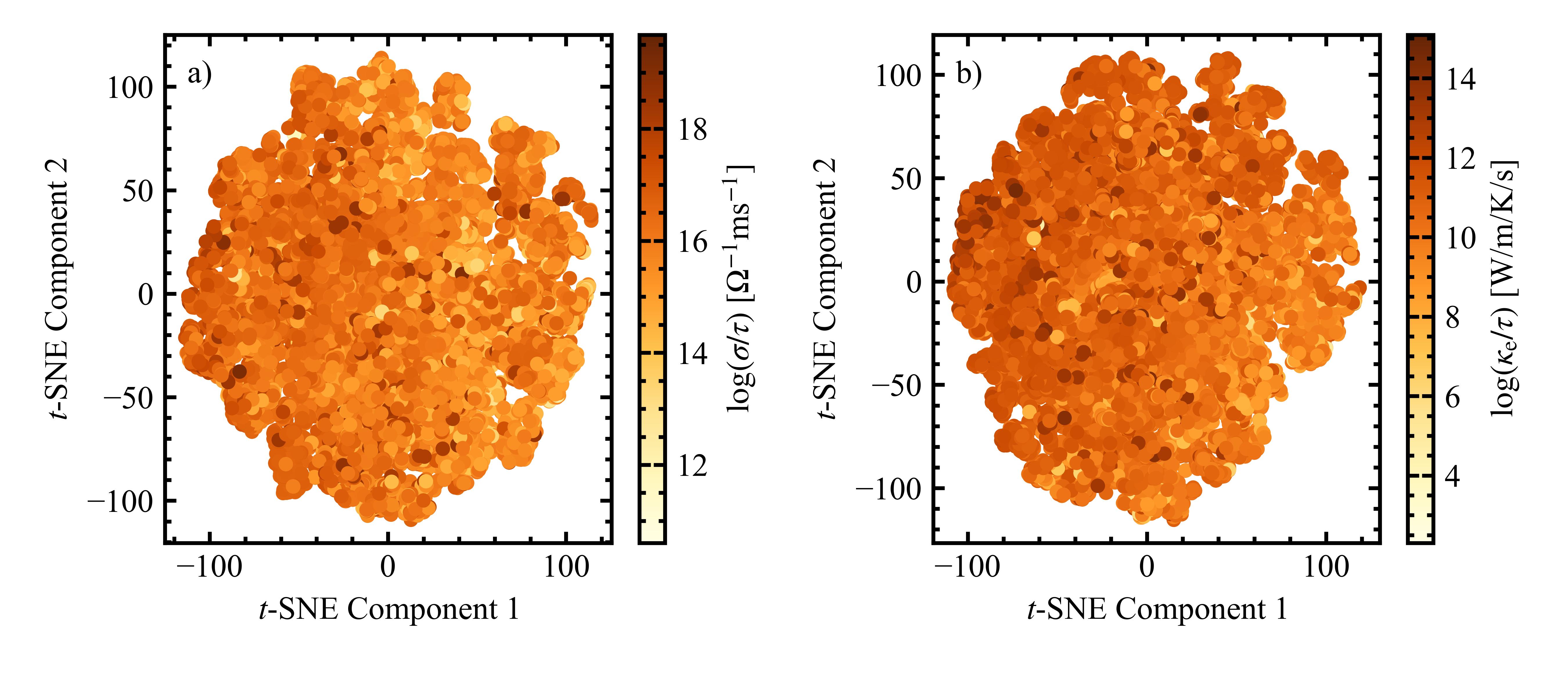}
    \caption{$t$-SNE visualisation. Each point is colour\textendash coded according to its value of (a) $\log(\sigma/\tau)$ and (b) $\log(\kappa_{\rm e}/\tau)$, respectively.}
    \label{fig:tsne-sigma-kappa}
\end{figure}

We extracted the final embedding representations of $\log(\sigma/\tau)$ and $\log(\kappa_{\rm e}/\tau)$ from TECSA-GNN and visualised them in a two-dimensional plane using $t$-SNE, as shown in Fig.~\ref{fig:tsne-sigma-kappa}. The visualised samples correspond to $T=300~\mathrm{K}$ and a carrier concentration of $n=10^{18}~\mathrm{cm}^{-3}$. 

\subsection{Global feature importance ranking for $\hat{S}$}
\label{sup:5}
In our case, we employ SHAP~\cite{lundberg2017unified} to quantify the influence of each component of the global feature vector $\mathbf{U}$ on the model's predictions. Within $\mathbf{U}$, the band gap $E_{\rm g}$ is supplied as prior knowledge, whereas all remaining dimensions are automatically constructed from the Magpie descriptors~\cite{ward2016general} based solely on the chemical formula. Magpie features are widely used in TE property prediction~\cite{vipin2024machine,al2024high,dunn2023machine}; each carries a clear physical meaning, yet they are rather elementary and therefore not immediately suitable for elucidating the connection between fundamental material properties and electronic transport performance. 

\setlength\LTcapwidth{\textwidth}
\begin{longtable}{clc}
\caption{Global feature importance ranked by mean absolute SHAP value.}
\label{tab:shap-features} \\
\toprule
Rank & Global feature & Mean-absolute SHAP value \\
\midrule
\endfirsthead
\multicolumn{3}{c}%
{{\tablename\ \thetable{} - continued from previous page}} \\
\toprule
Rank & Global feature & Mean-absolute SHAP value \\
\midrule
\endhead
\midrule \multicolumn{3}{r}{{Continued on next page}} \\
\endfoot
\botrule
\endlastfoot
1  & gap ($E_{\rm g}$)                                           & 0.014344 \\
2  & MagpieData range Electronegativity ($\Delta \chi$)          & 0.012311 \\
3  & frac p valence electrons ($f_{\rm p}$)                     & 0.012232 \\
4  & MagpieData range MendeleevNumber                             & 0.009984 \\
5  & MagpieData avg\_dev Row                                      & 0.008366 \\
6  & MagpieData mean SpaceGroupNumber                              & 0.008150 \\
7  & 3-norm                                                      & 0.007775 \\
8  & MagpieData minimum MendeleevNumber                           & 0.007489 \\
9  & MagpieData mean MendeleevNumber                              & 0.007297 \\
10 & MagpieData minimum Column                                     & 0.007127 \\
11 & MagpieData minimum Electronegativity                          & 0.007092 \\
12 & MagpieData range SpaceGroupNumber                             & 0.007067 \\
13 & MagpieData maximum NdValence                                   & 0.006795 \\
14 & MagpieData mode Row                                           & 0.006782 \\
15 & MagpieData range Column                                       & 0.006457 \\
16 & MagpieData avg\_dev NsUnfilled                                 & 0.005798 \\
17 & MagpieData minimum NpValence                                   & 0.005114 \\
18 & MagpieData avg\_dev NfValence                                  & 0.004988 \\
19 & MagpieData range MeltingT                                      & 0.004982 \\
20 & MagpieData mean GSvolume\_pa                                   & 0.004847 \\
\end{longtable}

In the main text, we provide physical interpretations for three representative features, namely the band gap ($E_{\rm g}$), the MagpieData range of electronegativity ($\Delta \chi$), and the fraction of $p$-valence electrons ($f_{\rm p}$). For reference, we list here the top twenty features ranked by their mean-absolute SHAP values (see Tab.~\ref{tab:shap-features}), while a more comprehensive table is available online~\cite{Zeng_TECSA-GNN}.

\subsection{Ablation studies and model efficiency assessment}
\label{sup:6}
The necessity of the model design and the appropriateness of the hyperparameter settings must be cross-validated through ablation studies. Here, we discuss the effects of varying the number of graph convolution layers, as well as the presence or absence of the global features $\mathbf{U}$ and the attention weights $\tilde{\alpha}_{ij}$ in the graph convolution process, on both model performance and computational efficiency.
\begin{figure}[h]
    \centering
    \includegraphics[width=\textwidth]{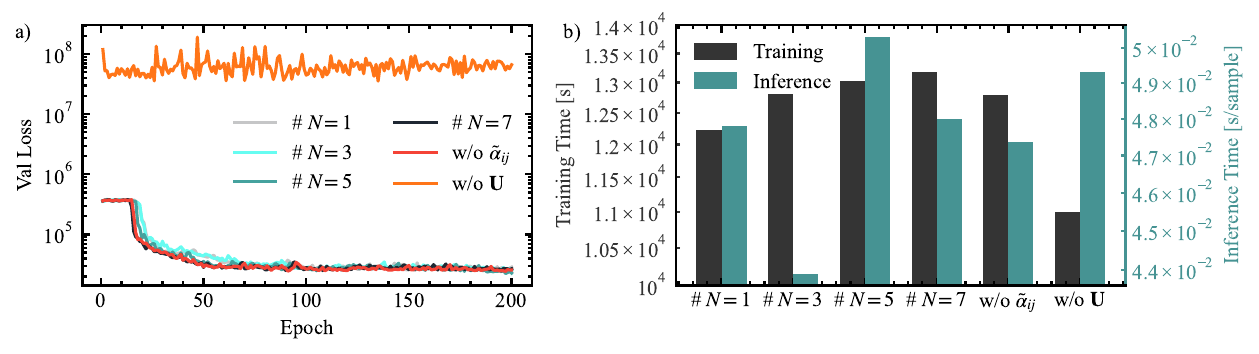}
    \caption{Ablation study results. (a) Test-set MSE loss curves under different configurations; all models were trained for 200 epochs. (b) Model efficiency evaluated in terms of training time (left $y$-axis) and inference time per sample (right $y$-axis).}
    \label{fig:abla-time}
\end{figure}

\begin{figure}[h]
    \centering
    \includegraphics[width=0.35\textwidth]{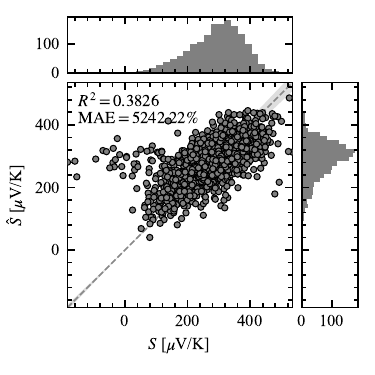}
    \caption{Parity plot of the TECSA-GNN (w/o $\mathbf{U}$) predicted $S$.}
    \label{fig:wo-u-s}
\end{figure}

For simplicity, we set the learning task to $p$-type $S$ at $T = 300~\mathrm{K}$ and $n = 10^{20}~\mathrm{cm^{-3}}$, using only 1,024 samples for the ablation study. In Fig.~\ref{fig:abla-time}, \# $N=5$ corresponds to the standard model used in the main text. In Fig.~\ref{fig:abla-time}(a), it is evident that removing $\mathbf{U}$ leads to a substantial drop in performance, indicating that $\mathbf{U}$ is indispensable for learning $S$. By contrast, within the limited number of training epochs, the standard model achieves near-optimal extrapolation, though differences with other configurations are minor. As shown in Fig.~\ref{fig:abla-time}(b), training time increases significantly with model complexity, whereas higher complexity does not guarantee superior accuracy. For example, the \# $N=7$ configuration requires the longest training time but achieves slightly lower accuracy than the standard setting. Interestingly, inference time per sample does not show a clear positive correlation with model complexity, though overall differences are small, possibly due to stochastic effects. Additionally, computing the attention weights increases training time by only $\sim 1.7\%$ while noticeably improving model performance. The performance of TECSA-GNN without $\mathbf{U}$ is shown in Fig.~\ref{fig:wo-u-s}.

The computations were conducted on a single node equipped with dual Intel Xeon E5-2640v4 Broadwell processors (2.4~GHz, 20 cores in total) and 128~GB of ECC DDR4 memory running at 2400~MHz. The node was fitted with four NVIDIA Tesla V100 Volta GPUs, each with 16~GB of memory and connected via NVLink. Nodes were interconnected using Intel Omni-Path Architecture (OPA) at 100~Gbps. Jobs were submitted via SLURM using one node, four tasks per node, and one GPU per task (\texttt{\#SBATCH -nodes=1}, \texttt{\#SBATCH -ntasks-per-node=4}, \texttt{\#SBATCH -gres=gpu:1}).

\subsection{Mitigating data imbalance during the fine-tuning process using a weighted-loss strategy}
\label{sup:fine-tuning}

\begin{figure}[h]
    \centering
    \includegraphics[width=\textwidth]{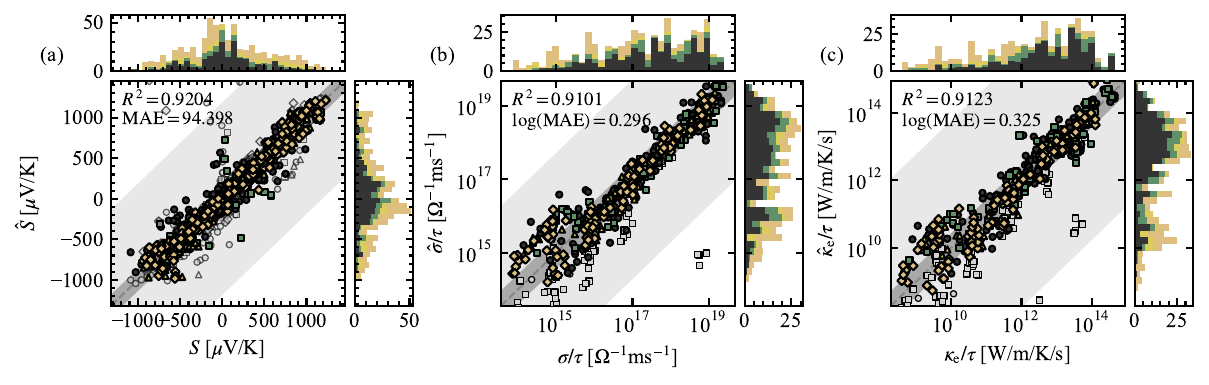}
    \caption{Parity plots of TECSA-GNN after fine-tuning on special structures. (a) $S$, (b) $\log(\sigma/\tau)$, and (c) $\log(\kappa_{\rm e}/\tau)$. Only test-set samples are shown. Circles, squares, triangles, and diamonds correspond to HH, FH, ZB, and RS structures. Grey markers denote predictions from the pretrained model, while coloured markers indicate predictions after fine-tuning. Light-grey bands represent one standard deviation of the prediction error of the pretrained model, with the corresponding range after fine-tuning shown in dark grey. Histograms along the top and right of each panel display the distributions of DFT reference values and model estimates, respectively.}
    \label{fig:fine-tune-supp}
\end{figure}

In the main text, we evaluated the generalization capability of TECSA-GNN on completely unseen structures and found that fine-tuning for only a small number of epochs can noticeably improve the model's generalization performance. However, a new issue that cannot be overlooked is that when the target-value distribution of new samples becomes imbalanced, the model often performs unsatisfactorily in sparsely populated regions (see Fig.~\ref{fig:fine-tune-supp}(b) and (c)). This phenomenon is quite common in machine-learning tasks.

To address this issue, an effective strategy is ``weighted-loss training''~\cite{steininger2021density}, in which higher weights are assigned to sparse samples in the loss function. More specifically, let $\mathcal{D}=\{(\mathcal{C}_i,y_i)\}_{i=1}^{N}$ denote the fine-tuning set for a given transport property, where $\mathcal{C}_i$ is the crystal input and $y_i$ is the corresponding DFT reference value. Following Ref.~\cite{steininger2021density}, we first estimate the empirical density of the target values using kernel density estimation (KDE)~\cite{chen2017tutorial}:
\begin{align}
p(y)=\frac{1}{Nh}\sum_{i=1}^{N}K\left(\frac{y-y_i}{h}\right),
\end{align}
where $K(\cdot)$ is the Gaussian kernel and $h$ is the bandwidth. The estimated density is then normalized as
\begin{align}
\tilde{p}(y)=\frac{p(y)-\min_{j}p(y_j)}{\max_{j}p(y_j)-\min_{j}p(y_j)} \in [0,1].
\end{align}

Accordingly, samples located in sparsely populated regions of the target space, particularly in the low-value regime highlighted in Fig.~\ref{fig:fine-tune-supp}, are assigned larger weights according to
\begin{align}
w(y_i)=
\frac{\max\left(1-\alpha \tilde{p}(y_i),\varepsilon\right)}
{\frac{1}{N}\sum_{j=1}^{N}\max\left(1-\alpha \tilde{p}(y_j),\varepsilon\right)},
\end{align}
where $\alpha \ge 0$ controls the strength of reweighting and $\varepsilon>0$ is a small constant introduced to prevent vanishing weights. By construction, the sample weights satisfy $\frac{1}{N}\sum_{i=1}^{N}w(y_i)=1$. The final weighted regression objective is defined as
\begin{align}
\mathcal{L}_{\mathrm{w}}=
\frac{1}{N}\sum_{i=1}^{N}w(y_i)\ell\left(f_{\theta}(\mathcal{C}_i),y_i\right),
\end{align}
where $f_{\theta}(\mathcal{C}_i)$ denotes the TECSA-GNN prediction and $\ell\left(\hat{y},y\right)$ is the base regression loss, taken here to be the mean-squared error. In this way, low-density samples contribute more strongly to the parameter updates during fine-tuning, thereby alleviating the tendency toward underfitting in the low-value region without modifying the model architecture.

\begin{figure}[h]
    \centering
    \includegraphics[width=.63\textwidth]{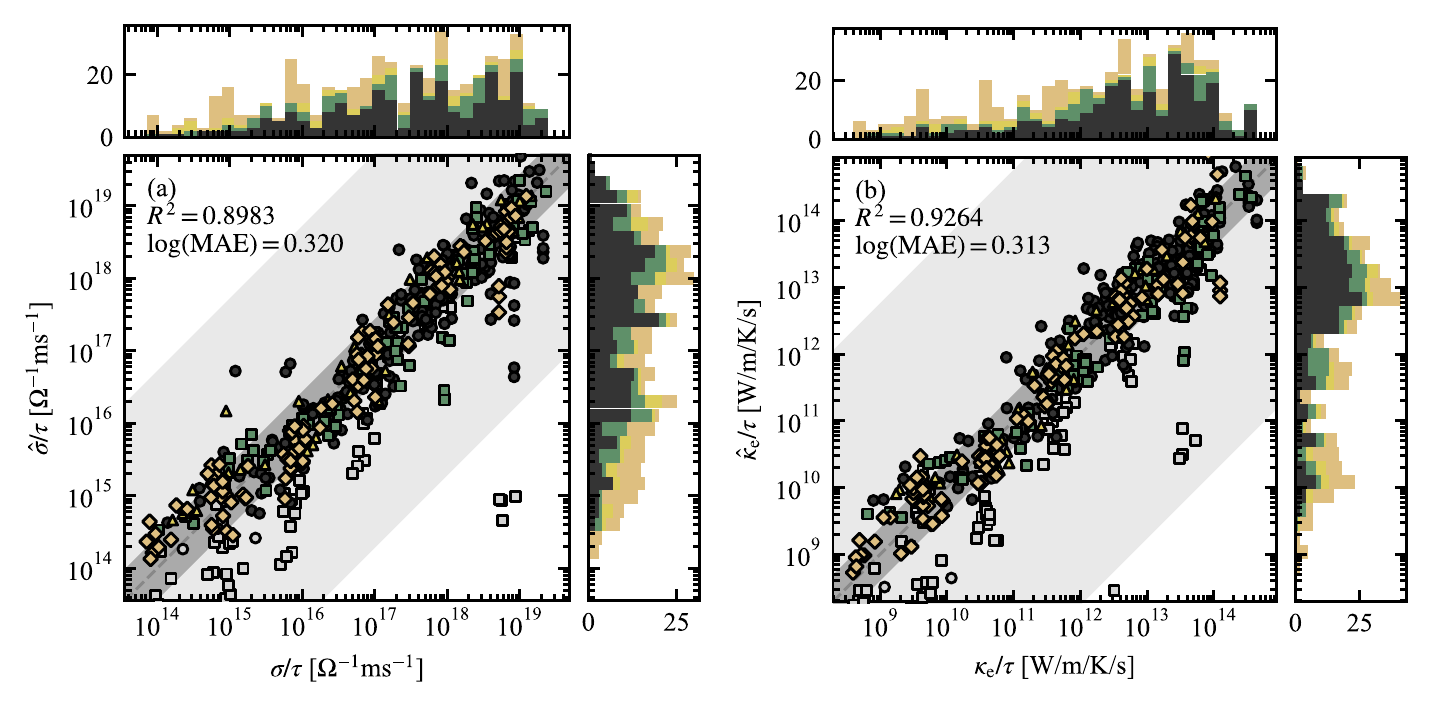}
    \caption{Parity plots of the fine-tuned model for special structures under the \textit{weighted-loss} training strategy. (a) $\log(\sigma/\tau)$, (b) $\log(\kappa_{\rm e}/\tau)$.}
    \label{fig:weighted-tuning}
\end{figure}

The fine-tuning results obtained with the weighted-loss training strategy are shown in Fig.~\ref{fig:weighted-tuning}. For $\sigma/\tau$ in Fig.~\ref{fig:weighted-tuning}(a), although more outliers appear in the high-value region, the overall errors remain within one standard deviation, while the low-value samples become more closely distributed around the reference line. The results for $\kappa_{\rm e}/\tau$ in Fig.~\ref{fig:weighted-tuning}(b) are even more encouraging. Although some accuracy is lost in the high-value region, the prediction accuracy for the low-value samples is significantly improved after fine-tuning.

\subsection{Details of the unsupervised learning method}

The purpose of employing $t$-SNE is to assess whether the learned embeddings of TECSA-GNN encode crystallographic similarity and doping symmetry, rather than to strictly partition materials into a predefined set of discrete categories. Methodologically, $t$-SNE is a nonlinear manifold learning technique that preserves neighborhood probabilities in the embedding space~\cite{maaten2008visualizing}. For high-dimensional representations $\mathbf{z}_i$ and $\mathbf{z}_j$, the pairwise similarity is defined as the conditional probability
\begin{align}
p_{j|i} =
\frac{\exp\left(-\|\mathbf{z}_i-\mathbf{z}_j\|^2/2\sigma_i^2\right)}
{\sum_{k\neq i}\exp\left(-\|\mathbf{z}_i-\mathbf{z}_k\|^2/2\sigma_i^2\right)}.
\end{align}

In the low-dimensional space, similarities are modeled using a Student-$t$ distribution,
\begin{align}
q_{ij} =
\frac{\left(1+\|\mathbf{y}_i-\mathbf{y}_j\|^2\right)^{-1}}
{\sum_{k\neq l}\left(1+\|\mathbf{y}_k-\mathbf{y}_l\|^2\right)^{-1}},
\end{align}
and the mapping is obtained by minimizing the Kullback-Leibler (KL) divergence~\cite{joyce2025kullback},
\begin{align}
\mathcal{L}=
\sum_{i\neq j} p_{ij}\log\frac{p_{ij}}{q_{ij}},
\end{align}
thereby preserving local neighborhood structures. Intuitively, $t$-SNE emphasizes which samples are close to one another in the high-dimensional embedding space and maintains this proximity in the two-dimensional projection. This makes it particularly well suited for examining whether the learned representation forms continuous regions, whether structurally related crystal systems appear adjacent, and whether symmetric patterns associated with doping types naturally emerge.

In contrast, $k$-means is a partition-based clustering algorithm with the objective
\begin{align}
\mathcal{L}=\sum_{k=1}^{K}\sum_{\mathbf{z}_i\in C_k}\|\mathbf{z}_i-\boldsymbol{\mu}_k\|^2,
\end{align}
where $\boldsymbol{\mu}_k$ denotes the centroid of cluster $k$. It produces hard assignments by minimizing within-cluster variance. In this sense, $k$-means answers the question of how to partition the samples, whereas $t$-SNE probes whether the geometry of the embedding space itself carries physically meaningful structure. Since our goal is to diagnose representation quality rather than enforce categorical separation, $t$-SNE serves as the more appropriate visualization tool in this context.

\begin{figure}[t]
    \centering
    \includegraphics[width=.9\textwidth]{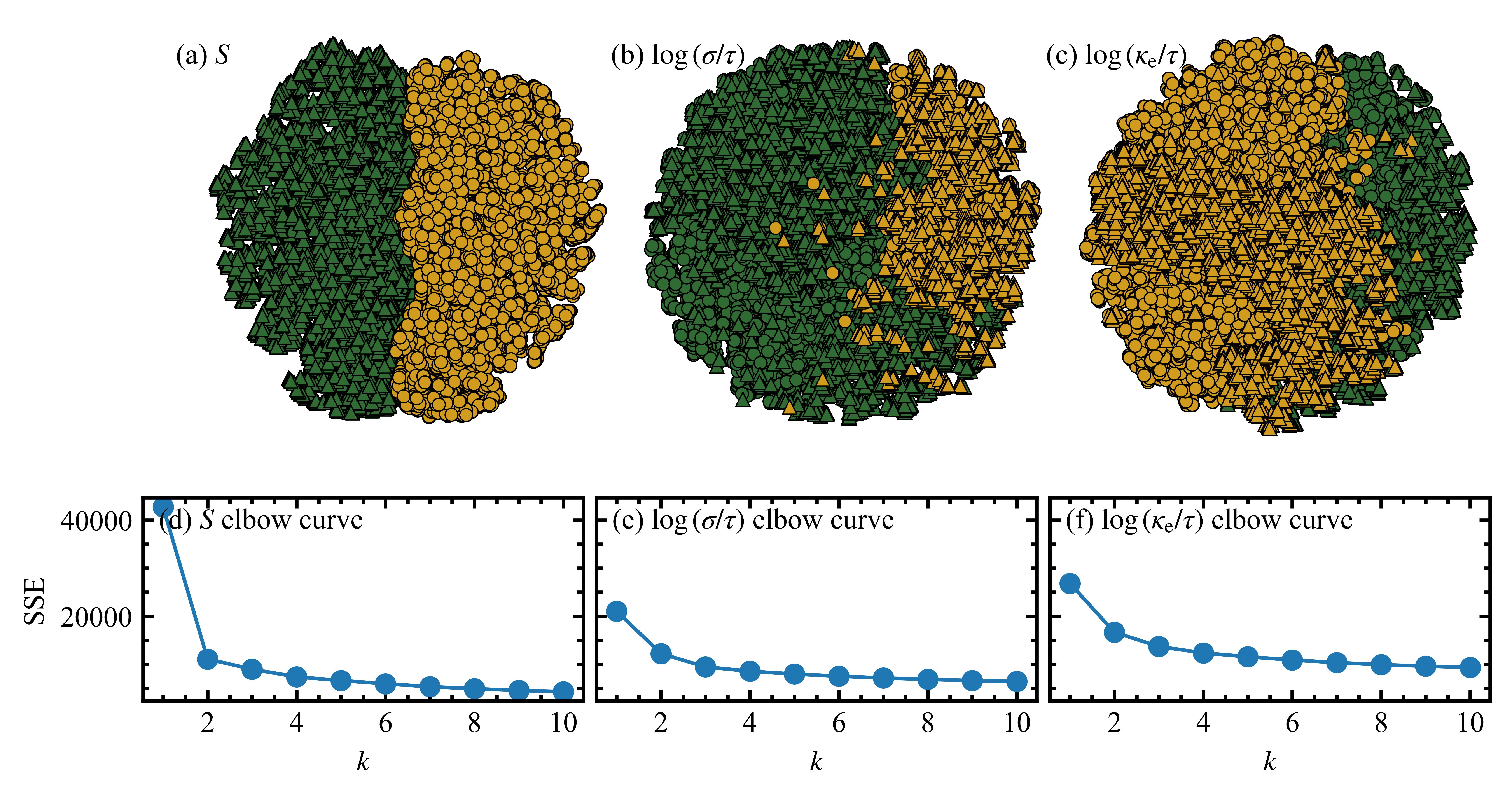}
    \caption{$k$-means clustering and its $t$-SNE visualization. (a) $S$, (b) $\log\left(\sigma/\tau\right)$, and (c) $\log\left(\kappa_{\rm e}/\tau\right)$ at 300~K and $n=10^{18}$~cm$^{-3}$. (d)-(f) show the corresponding $k$-means elbow curves. Here we choose $k=2$. Circular markers denote $n$-type samples, while triangular markers denote $p$-type samples.}
    \label{fig:kmeans-k2}
\end{figure}

We next extract the sample embeddings from the pretrained models for $S$, $\log\left(\sigma/\tau\right)$, and $\log\left(\kappa_{\rm e}/\tau\right)$, respectively, and perform $k$-means clustering for $t$-SNE visualization, as shown in Fig.~\ref{fig:kmeans-k2}.

In practice, the optimal number of clusters $k$ in $k$-means is commonly identified using the elbow method~\cite{nainggolan2019improved}, in which $k$ is selected at the point where the sum of squared errors (SSE) exhibits a pronounced inflection. As shown in Fig.~\ref{fig:kmeans-k2}(d)-(f), this criterion leads to $k=2$ in all three cases.

\begin{figure}[t]
    \centering
    \includegraphics[width=.9\textwidth]{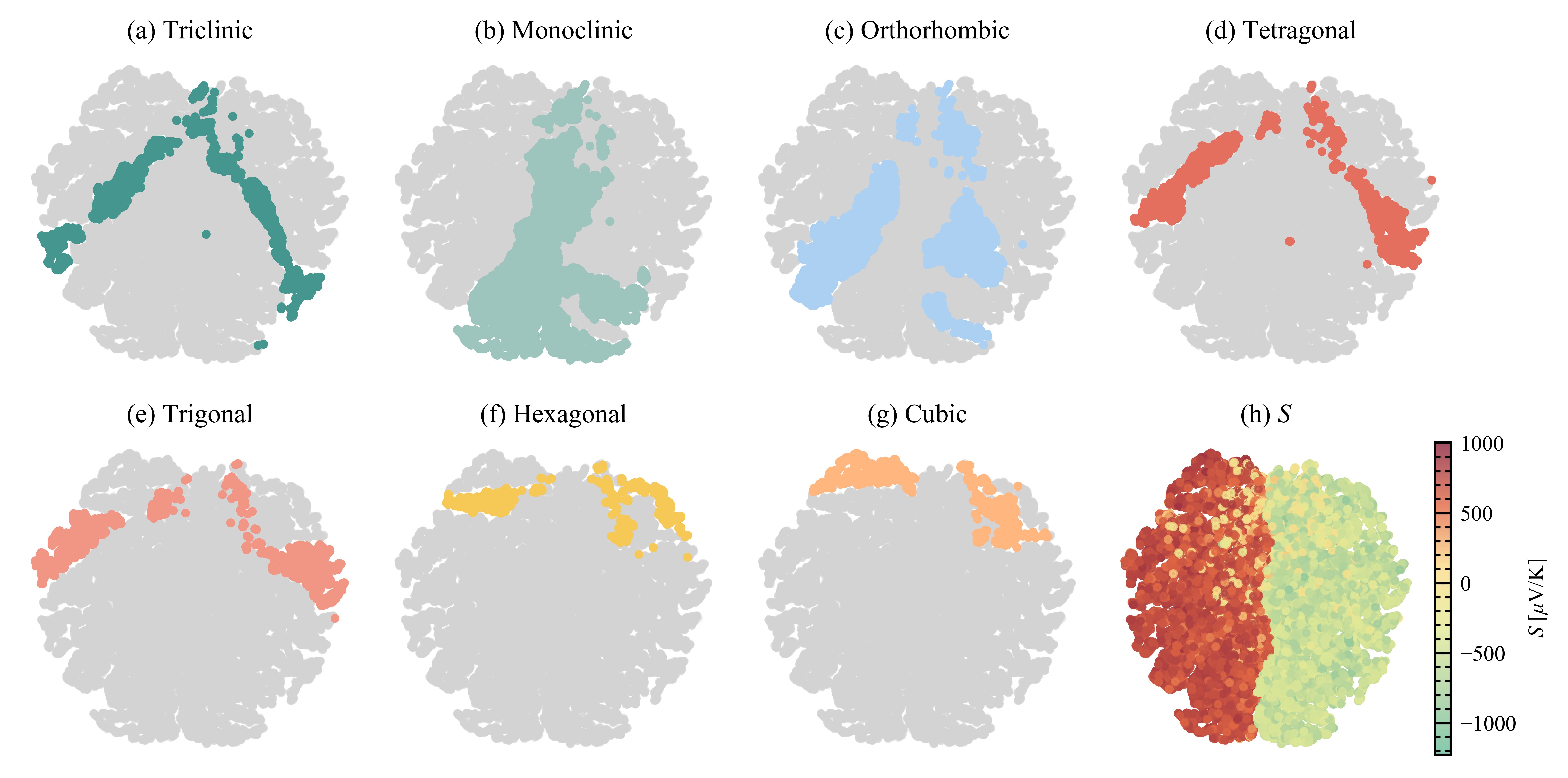}
    \caption{$t$-SNE visualization of sample embeddings across different crystal systems. (a)-(g) Each colour denotes one of the seven crystal systems in the two-dimensional $t$-SNE plane, with grey points indicating samples from other systems. (h) Colour mapping of the Seebeck coefficient $S$ illustrates both its magnitude and spatial distribution.}
    \label{fig:tsne-cs-s-supp}
\end{figure}

We observe in Fig.~\ref{fig:kmeans-k2}(a) that the $p$- and $n$-type samples are cleanly separated into left and right regions. This is consistent with our earlier conjecture in the main text that the $x$-component of the $t$-SNE projection in Fig.~\ref{fig:tsne-cs-s-supp} largely encodes the doping-type label. The effect is particularly pronounced for the embeddings associated with $S$. A plausible explanation is that the doping type almost entirely determines the sign of $S$, with exceptions arising only in regimes such as low carrier concentrations where bipolar transport becomes relevant~\cite{park2021optimal}. By contrast, for $\log\left(\sigma/\tau\right)$ and $\log\left(\kappa_{\rm e}/\tau\right)$, the $k$-means clustering does not yield an equally clear physical interpretation, suggesting that the learned embeddings for these quantities do not organize primarily along doping symmetry.

\begin{figure}[h]
    \centering
    \includegraphics[width=.9\textwidth]{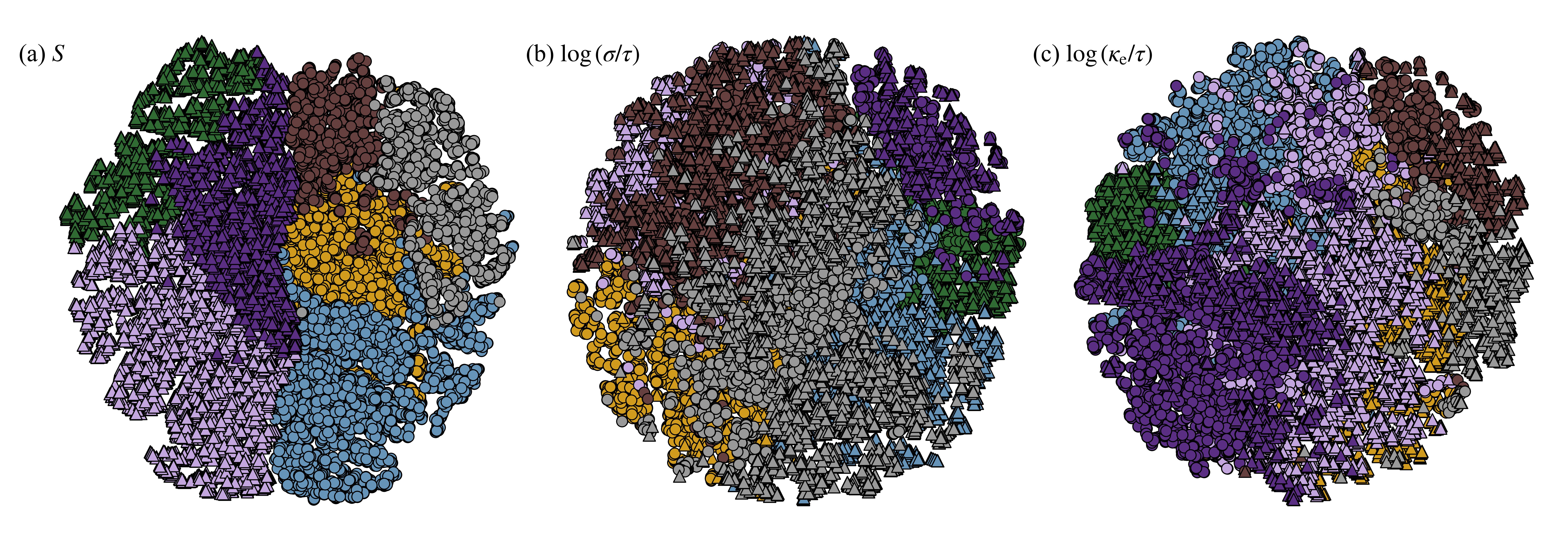}
    \caption{$k$-means ($k=7$) clustering for embeddings. (a) $S$, (b) $\log\left(\sigma/\tau\right)$, and (c) $\log\left(\kappa_{\rm e}/\tau\right)$.}
    \label{fig:k7}
\end{figure}

If we instead impose prior knowledge of the seven crystal systems and set $k=7$, the resulting clustering is shown in Fig.~\ref{fig:k7}. In Fig.~\ref{fig:k7}(a), the inter-cluster separation appears relatively distinct; however, the identified clusters do not align well with the true crystallographic groupings displayed in Fig.~\ref{fig:tsne-cs-s-supp}. In Figs.~\ref{fig:k7}(b) and (c), the cluster distributions become noticeably mixed, and a clear physical interpretation is no longer evident.

In summary, the role of $t$-SNE is to reveal whether the geometry of the embedding space carries physically meaningful structure, rather than to forcibly partition samples into a fixed number of categories. By contrast, $k$-means is more naturally suited to tasks where class labels are either known or expected to exist, such as materials classification~\cite{wu2025hierarchy} or cell image categorization~\cite{zhou2025imaging}.

\subsection{Electronic transport coefficients calculated within the electron-phonon approximation: the case of LiMgSb}

To go beyond the constant relaxation time approximation (CRTA, $\tau=10^{-14}$ s), we further considered electron-phonon approximation (EPA)~\cite{wilson2020parametric}, which explicitly accounts for carrier scattering by lattice vibrations and therefore provides a more realistic description of electrical transport than a fixed relaxation time. In the present context, the EPA calculation is introduced not to replace the CRTA-based analysis adopted throughout the manuscript, but rather to provide a higher-fidelity first-principles validation for a representative material system.

\begin{figure}[h]
    \centering
    \includegraphics[width=\textwidth]{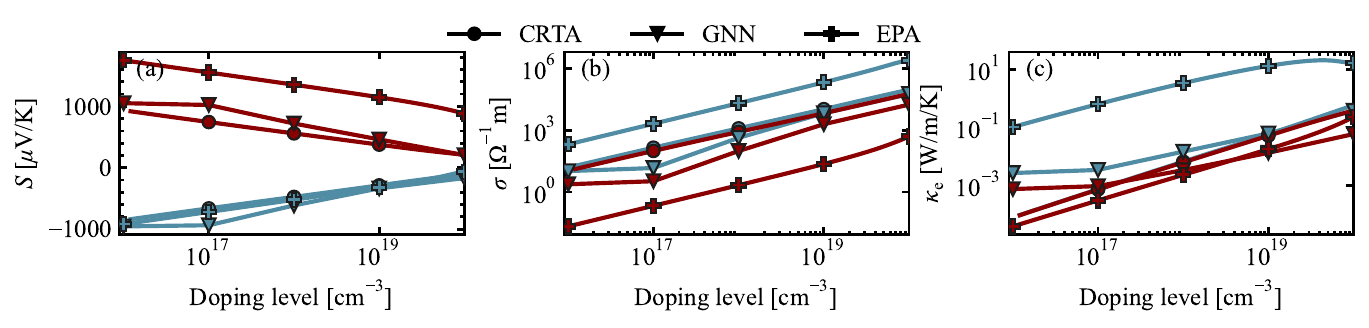}
    \caption{Comparison of the electronic transport coefficients of LiMgSb obtained from the CRTA-based DFT calculations, TECSA-GNN estimations, and EPA-based DFT calculations. (a)-(c) $S$, $\sigma$, and $\kappa_{\rm e}$ as functions of carrier concentration. Blue and red curves denote $n$-type and $p$-type doping, respectively.}
    \label{fig:epc}
\end{figure}

In our case, calculations were carried out using the {\sc Quantum} ESPRESSO package~\cite{Giannozzi_2017,Giannozzi_2009} with the PBEsol exchange-correlation functional~\cite{PhysRevLett.100.136406}. Optimized norm-conserving Vanderbilt pseudopotentials from the PseudoDojo library~\cite{van2018pseudodojo} were employed. Structural relaxations were performed until the residual force on each atom was below 0.01~eV/\AA, using a plane-wave cutoff energy of 80~Ry. The Brillouin zone was sampled by a Monkhorst-Pack $\mathbf{k}$-point mesh with a spacing of approximately 0.15 \AA$^{-1}$.

Phonon and EPA calculations were then performed within the QE-PHOEBE workflow~\cite{Cepellotti_2022}. In practice, a $6\times6\times6$ mesh was adopted for both the electronic $\mathbf{k}$-points in \texttt{pw.x} and the phonon $q$-points in \texttt{ph.x}, following the requirement of the PHOEBE implementation. The electronic Boltzmann transport equation was solved using the adaptive smearing scheme, and convergence with respect to a dense $48\times48\times48$ $\mathbf{k}$-point mesh was confirmed.

As illustrated in Fig.~\ref{fig:epc}, the EPA calculations preserve the same overall carrier-concentration dependence as the CRTA-based results and the GNN predictions, namely, a reduction in $|S|$ together with increasing $\sigma$ and $\kappa_{\mathrm e}$ as the carrier concentration increases. However, noticeable quantitative discrepancies are observed in the transport coefficients once electron-phonon scattering is treated explicitly. In particular, the absolute magnitudes of $\sigma$ and $\kappa_{\mathrm e}$ differ substantially between the EPA and CRTA results, and the deviation is strongly carrier-type dependent. Moreover, for $p$-type LiMgSb, the Seebeck coefficient $S$ obtained from EPA is also significantly larger than that from the CRTA-based calculation over the investigated doping range.

These results indicate that the CRTA-based treatment remains useful for capturing the dominant qualitative transport trends and for maintaining consistency with the database label space used for model training, but it is insufficient for fully quantitative transport predictions for specific materials. In this sense, the EPA calculation presented here serves as a higher-fidelity first-principles validation for a representative case, while also highlighting the intrinsic limitation of adopting a fixed relaxation time for different carrier types and transport channels.

Accordingly, we have revised the manuscript and the Supplementary Information to clarify the role of the fixed relaxation time in the training dataset and to explicitly distinguish the CRTA-based screening analysis from the EPA-based higher-fidelity validation.

\subsection{High-fidelity band-structure calculations}

\begin{figure}[h]
    \centering
    \includegraphics[width=.45\textwidth]{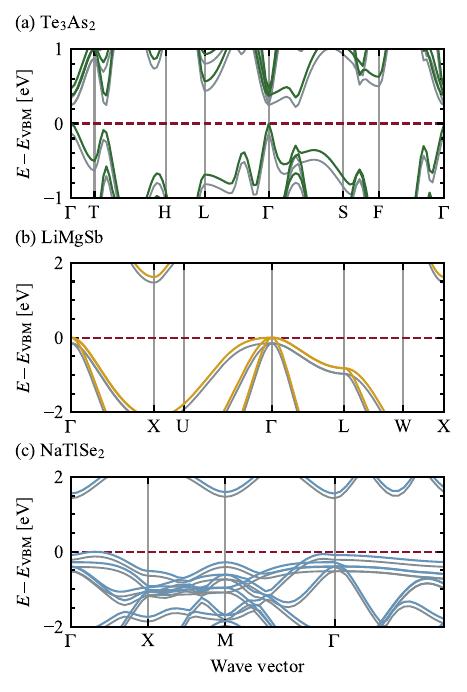}
    \caption{Band structures of (a) Te$_3$As$_2$, (b) LiMgSb, and (c) NaTlSe$_2$ shown using two different energy references. The gray lines correspond to the original DFT-derived alignment referenced to $E_{\rm Fermi}$, whereas the lines in the other colors are aligned to the VBM. In the VBM-aligned representation, the zero-energy line corresponds to the VBM, and the relative position of the original DFT-derived Fermi level can be directly compared with the band edges.}
    \label{fig:new-bands}
\end{figure}

\emph{Using the VBM as the zero-energy reference.---} For semiconducting systems, the absolute position of the Fermi level obtained from standard DFT calculations within the band gap is usually not the most appropriate reference for discussing the frontier electronic structure, since it may be sensitive to numerical details such as electron filling and $\mathbf{k}$-point sampling. Therefore, when focusing on band-edge features relevant to transport, it is more appropriate to align the energy reference to the VBM. This treatment has been widely adopted in the literature; for example, in Ref.~\cite{PhysRevB.95.165204}, the Fermi level is explicitly set to the top of the valence bands for semiconducting CoSb$_3$.

Following this convention, we re-examined the band structures of Te$_3$As$_2$, LiMgSb, and NaTlSe$_2$ using the VBM as the common energy reference, and compared them with the corresponding results referenced to the original DFT-derived Fermi level. The comparison shows that, for all three systems, the DFT-derived Fermi level lies inside a finite gap rather than crossing dispersive bands. In other words, when the band structures are aligned to the VBM, the original DFT-derived Fermi level is located below the zero-energy reference. 

Under this VBM-aligned description, the main qualitative conclusions of the original discussion remain valid, although some wording should be made more precise. In particular, Te$_3$As$_2$ is more appropriately described as a narrow-gap semiconductor with relatively dispersive band-edge states, rather than as a nearly metallic system. Nevertheless, its relatively strong band-edge dispersion remains evident, and thus the conclusion that it exhibits a comparatively stronger electronic transport response is still justified. For LiMgSb and NaTlSe$_2$, the relative band-edge characteristics are also preserved: NaTlSe$_2$ still shows flatter valence-band-edge states, corresponding to a larger effective mass, which is favorable for a higher $| S \vert $ but unfavorable for $\kappa_{\rm e}$, whereas LiMgSb exhibits more dispersive band-edge states and therefore a more balanced transport behavior. This trend is consistent with the general thermoelectric understanding that flatter bands tend to enhance thermopower, while more dispersive bands favor carrier mobility and $\kappa_{\rm e}$.

\emph{High-fidelity HSE functional.---} The band-structure calculations presented in the main text were performed using the Perdew-Burke-Ernzerhof (PBE) functional~\cite{PhysRevLett.77.3865}, for which the exchange-correlation energy is defined as
\begin{align}
    E_{{\rm xc}}^{{\rm PBE}}
= \int \Big( \epsilon_x^{{\rm PBE}}\big(n(\boldsymbol{r}),\nabla n(\boldsymbol{r})\big)
         + \epsilon_c^{{\rm PBE}}\big(n(\boldsymbol{r}),\nabla n(\boldsymbol{r})\big) \Big)\, n(\boldsymbol{r})\, d\boldsymbol{r},
\end{align}
where \(n(\boldsymbol{r})\) is the electron density at position \(\boldsymbol{r}\), \(\epsilon_x^{{\rm PBE}}\) is the exchange-energy density, and \(\epsilon_c^{{\rm PBE}}\) is the correlation-energy density. The PBE functional represents a practical compromise between computational accuracy and efficiency. Although it is widely used, because it depends only on the electron density and its gradient, its ability to describe strongly correlated materials, especially those containing localized \(d/f\) electrons, remains limited~\cite{chen2025ecd}. Here, for the calculations shown in Fig.~\ref{fig:new-bands}, we further adopted the Heyd-Scuseria-Ernzerhof (HSE) functional~\cite{hse} to improve the fidelity. This functional incorporates short-range exact Hartree-Fock exchange together with a screened Coulomb potential, thereby providing higher fidelity at the cost of greater computational expense. The HSE exchange-correlation energy is written as
\begin{align}
    E_{xc}^{{\rm HSE}} = \alpha E_{x}^{{\rm HF,SR}}(\omega)+ (1-\alpha) E_{x}^{{\rm PBE,SR}}(\omega)+ E_{x}^{{\rm PBE,LR}}(\omega)+ E_{c}^{{\rm PBE}},
\end{align}
where \(E_{x}^{\text{HF,SR}}(\omega)\) denotes the short-range Hartree-Fock exchange energy, \(E_{x}^{\text{PBE,SR}}(\omega)\) and \(E_{x}^{\text{PBE,LR}}(\omega)\) denote the short-range and long-range PBE exchange energies, respectively, and \(E_{c}^{\text{PBE}}\) is the PBE correlation energy. The parameter \(\alpha\) is the mixing coefficient, which is commonly taken as 0.25, and \(\omega\) is the screening parameter.

To reduce the computational cost, the calculation in Fig.~\ref{fig:new-bands}(c) was carried out using the primitive cell, which leads to some change in the band shape; by contrast, no obvious change in the band shape is observed in Fig.~\ref{fig:new-bands}(a) and (b). In fact, for the present cases, the band structures obtained with the PBE functional are already sufficient to support the analysis of the physical origin of the electronic transport behavior.

In summary, adopting the VBM as the energy reference does not change the relative trends in the electronic transport behavior of these materials, but provides a more physically meaningful description of their band-edge electronic structures and avoids over-interpreting the absolute position of the raw DFT-derived Fermi level within the band gap.

\subsection{Evaluation of representative MLIP-oriented GNN models for the direct prediction of electronic transport coefficients}

The baseline models presented in the main-text benchmark were all developed for direct crystal-property prediction. In recent years, however, many GNN models designed specifically for training machine-learning interatomic potentials (MLIPs) have also emerged, such as CHGNet~\cite{chgnet} and M3GNet~\cite{m3gnet}. By replacing the final regression head with a scalar-output predictor, the architectures of CHGNet and M3GNet can likewise be adapted for learning electronic transport coefficients.

\emph{M3GNet.---} M3GNet is a many-body GNN originally developed for universal MLIP learning in crystalline materials. For a crystal specified by atomic numbers $\{Z_i\}$, atomic Cartesian coordinates $\{\mathbf{r}_i\}$, and lattice matrix $\mathbf{L}$, the structure is represented as a material graph in which atom $i$ is encoded through an embedding of $Z_i$, whereas bond $(i,j)$ is encoded from the interatomic distance $r_{ij}=\|\mathbf{r}_i-\mathbf{r}_j\|$ via a radial basis expansion. To account for local many-body geometry, M3GNet further constructs triplets $(j,i,k)$ and evaluates the associated bond angle $\theta_{jik}$, such that each bond representation depends not only on the bonded pair itself but also on its angular environment. Denoting the initial atom and bond features by $\mathbf{v}_i^{(0)}$ and $\mathbf{e}_{ij}^{(0)}$, respectively, the input encoding can be written as
\begin{align}
\mathbf{v}_i^{(0)} &= \mathrm{Emb}(Z_i),\\
\mathbf{e}_{ij}^{(0)} &= \phi_r(r_{ij}),
\end{align}
where $\mathrm{Emb}(\cdot)$ denotes the learnable atomic embedding and $\phi_r(\cdot)$ the radial basis expansion of the interatomic distance.

Within each message-passing block, the bond feature is first refined by aggregating the local three-body information associated with triplets centred at atom $i$, after which the updated bond feature is used to propagate messages over the material graph. In compact form, this procedure is written as
\begin{align}
\tilde{\mathbf{e}}_{ij}^{(t)} &= \Psi_{\mathrm{mb}} \left(\mathbf{e}_{ij}^{(t)},\{\mathbf{e}_{ik}^{(t)},\theta_{jik}\}_{k\in\mathcal{N}(i)}\right),\\
\mathbf{v}_i^{(t+1)} &= \Psi_{\mathrm{v}} \left(\mathbf{v}_i^{(t)},\{\tilde{\mathbf{e}}_{ij}^{(t)}\}_{j\in\mathcal{N}(i)}\right),
\end{align}
where $\mathcal{N}(i)$ is the neighbour set of atom $i$, $\Psi_{\mathrm{mb}}$ denotes the many-body update from angular environments to bond features, and $\Psi_{\mathrm{v}}$ the subsequent graph-based node update. After $T$ blocks, the crystal-level representation $\mathbf{h}_{\mathrm{crystal}}$ is obtained through a graph-level readout over the final node features,
\begin{align}
\mathbf{h}_{\mathrm{crystal}}=\mathrm{Readout} \left(\{\mathbf{v}_i^{(T)}\}\right).
\end{align}

In its original formulation, M3GNet predicts the total energy as a sum of atom-wise contributions, $E=\sum_i E_i$, from which the atomic forces and virial stress are obtained by automatic differentiation with respect to atomic coordinates and strain, $\mathbf{F}_i=-\frac{\partial E}{\partial \mathbf{r}_i}$ and $\boldsymbol{\sigma}=\frac{1}{V}\frac{\partial E}{\partial \boldsymbol{\varepsilon}}$, with $V$ being the unit-cell volume and $\boldsymbol{\varepsilon}$ the strain tensor. In the present work, we retain the M3GNet backbone as a crystal encoder but replace the original energy head with a scalar regression module for direct prediction of an electronic transport coefficient $y$:
\begin{align}
\hat{y}=f_{\mathrm{head}} \left(\mathbf{h}_{\mathrm{crystal}}\right),
\end{align}
where $f_{\mathrm{head}}$ is the final multilayer perceptron. In this manner, the many-body structural descriptors learned by M3GNet are repurposed from interatomic-potential modelling to crystal-level electronic transport prediction.

\emph{CHGNet.---} Distinct from conventional crystal graph models, CHGNet represents a structure using both an atom graph and an auxiliary bond graph, thereby enabling atom, bond, and angle features to be updated in a coupled manner throughout the network. For a crystal specified by atomic numbers $\{Z_i\}$ and atomic Cartesian coordinates $\{\mathbf{r}_i\}$, neighbouring atoms within a cutoff radius define the edges of the atom graph, while a triplet $(i,j,k)$ defines an angular interaction between the two connected bonds $(i,j)$ and $(j,k)$ in the bond graph. Based on basis expansions of interatomic distances and bond angles, CHGNet initializes atom features $\mathbf{v}_i^{(t)}$, bond features $\mathbf{e}_{ij}^{(t)}$, and angle features $\mathbf{a}_{ijk}^{(t)}$, which are then updated jointly through coupled atom, bond, and angle convolutions as
\begin{align}
\mathbf{v}_i^{(t+1)}&=\mathbf{v}_i^{(t)}+L_v^{(t)} \left[\sum_j \mathbf{e}_{ij}^{a}\odot \phi_v^{(t)} \left(\mathbf{v}_i^{(t)} \Vert \mathbf{v}_j^{(t)} \Vert \mathbf{e}_{ij}^{(t)}\right)\right],\\
\mathbf{e}_{jk}^{(t+1)}&=\mathbf{e}_{jk}^{(t)}+L_e^{(t)} \left[\sum_i \mathbf{e}_{ij}^{b}\odot \mathbf{e}_{jk}^{b}\odot \phi_e^{(t)} \left(\mathbf{e}_{ij}^{(t)} \Vert \mathbf{e}_{jk}^{(t)} \Vert \mathbf{a}_{ijk}^{(t)} \Vert \mathbf{v}_j^{(t+1)}\right)\right],\\
\mathbf{a}_{ijk}^{(t+1)}&=\mathbf{a}_{ijk}^{(t)}+\phi_a^{(t)} \left(\mathbf{e}_{ij}^{(t+1)} \Vert \mathbf{e}_{jk}^{(t+1)} \Vert \mathbf{a}_{ijk}^{(t)} \Vert \mathbf{v}_j^{(t+1)}\right),
\end{align}
where $\Vert$ denotes feature concatenation, $\odot$ element-wise multiplication, $L_v^{(t)}$ and $L_e^{(t)}$ linear layers, $\phi_v^{(t)}$, $\phi_e^{(t)}$, and $\phi_a^{(t)}$ gated multilayer perceptrons, and $\mathbf{e}_{ij}^{a}$ and $\mathbf{e}_{ij}^{b}$ distance-dependent weights introduced to ensure smooth message passing near the graph cutoffs. A defining feature of CHGNet is that the intermediate atom features are explicitly regularized by site-resolved magnetic moments, such that the latent representation used for downstream prediction is constrained by charge-state-related information; in the original model, the magmom of atom $i$ is predicted after three interaction blocks as
\begin{align}
m_i=L_m \left(\mathbf{v}_i^{(3)}\right).
\end{align}

After this charge-constrained stage, the final atom features are used to predict the target property. In the original CHGNet formulation, the model outputs the total energy and subsequently obtains forces and stress by automatic differentiation with respect to atomic coordinates and strain. In the present work, we retain the CHGNet backbone as a crystal encoder but replace the original energy output module with a scalar regression head for direct prediction of an electronic transport coefficient $y$, namely $\hat{y}=f_{\mathrm{head}}(\mathbf{h}_{\mathrm{crystal}})$, where $\mathbf{h}_{\mathrm{crystal}}$ denotes the graph-level representation generated by the CHGNet backbone and $f_{\mathrm{head}}$ is the final multilayer perceptron. In this manner, the coupled atom--bond--angle representations and the charge-informed latent features learned by CHGNet are repurposed for crystal-level electronic transport prediction.

\begin{figure}[t]
    \centering
    \includegraphics[width=\textwidth]{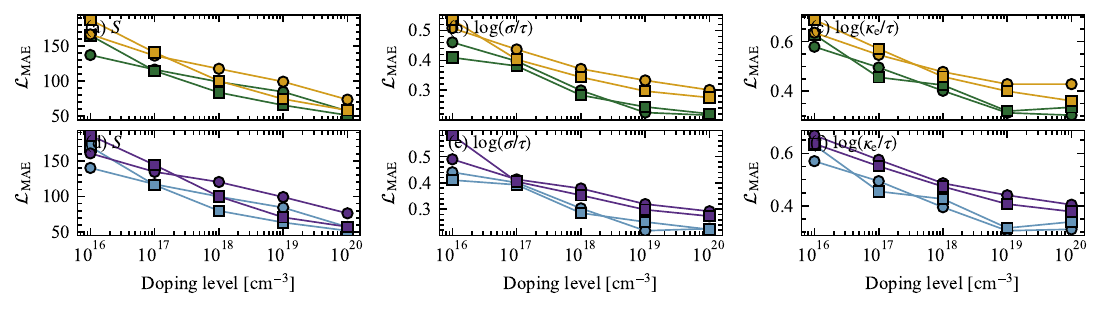}
    \caption{Error evaluation of representative MLIP-based GNN models on electronic transport coefficients. (a)-(c) correspond to CHGNet, and (d)-(f) correspond to M3GNet. Circular markers denote $n$-type samples and square markers denote $p$-type samples; green and blue indicate the corresponding training set results, while yellow and purple indicate the test set.}
    \label{fig:mlip-gnn}
\end{figure}

\emph{Performance on electronic transport prediction.---}
In practice, we attempted to use M3GNet and CHGNet to learn $\left\{S,\log\left(\sigma/\tau\right),\log\left(\kappa_{\rm e}/\tau\right)\right\}$ at a fixed temperature of $T=300~{\rm K}$, and the corresponding results are shown in Fig.~\ref{fig:mlip-gnn}. Qualitatively, the TECSA-GNN proposed in this work shows clearly better predictive reliability for $S$ than M3GNet and CHGNet. Interestingly, for the task of $\log\left(\sigma/\tau\right)$, when the doping level $n$ lies in the range $\left[10^{18},10^{20}\right]~\mathrm{cm}^{-3}$, M3GNet and CHGNet in fact exhibit slightly lower errors and outperform the other mentioned GNN models for direct crystal-property estimation.

This behaviour is physically reasonable. The Seebeck coefficient $S$ is highly sensitive to the fine electronic-structure asymmetry around the Fermi level, including band-edge curvature, carrier polarity, and the competition between localized and dispersive states. In this respect, the superior performance of TECSA-GNN is consistent with its multiscale design, which explicitly couples global physicochemical descriptors with atomic, bond, and angular representations and is therefore better suited to learning the intrinsic structure-transport relations underlying $S$. By contrast, both M3GNet and CHGNet were originally developed as MLIP backbones rather than direct transport predictors. Their latent representations are primarily optimized for encoding local geometric environments and, in the case of CHGNet, charge-state-related information~\cite{chgnet,m3gnet}. Such inductive biases can still be beneficial after replacing the original energy head with a scalar regressor, especially in the high-doping regime where $\log(\sigma/\tau)$ is more directly associated with the overall conductive character and local bonding environment of the crystal. Moreover, since $\sigma/\tau$ is defined under the constant-relaxation-time approximation, its label itself contains an additional level of physical simplification, which may further narrow the performance gap among different backbones in selected concentration windows. Therefore, the local advantage of M3GNet and CHGNet at high carrier concentrations should be understood as a limited transfer benefit of MLIP-oriented representations, rather than evidence against the overall superiority of TECSA-GNN for direct electronic transport prediction.

\subsection{Complete set of candidate materials in high-throughput screening}
\label{sup:4}

\begin{figure}[h]
    \centering
    \includegraphics[width=.7\textwidth]{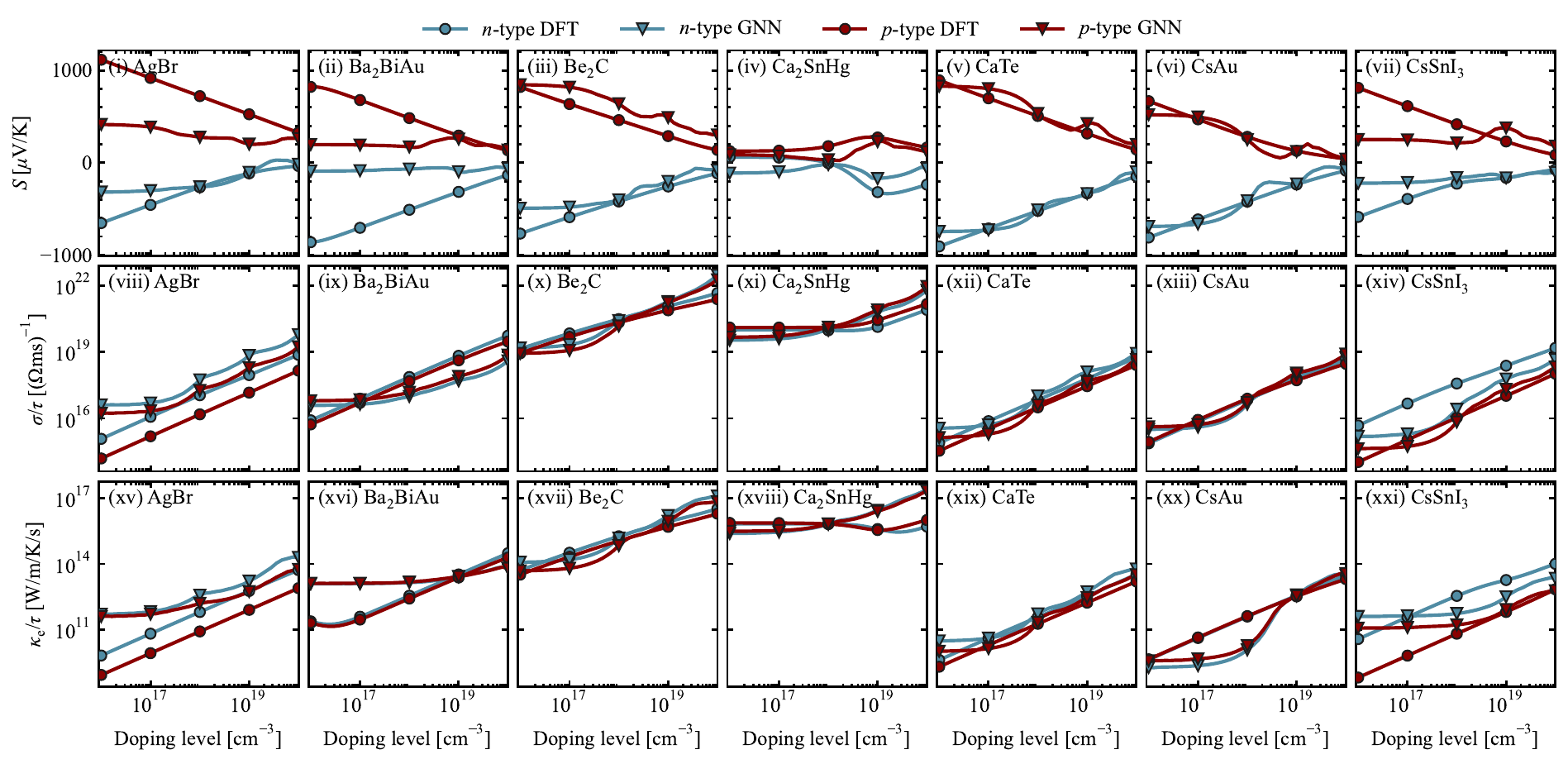}
    \includegraphics[width=.7\textwidth]{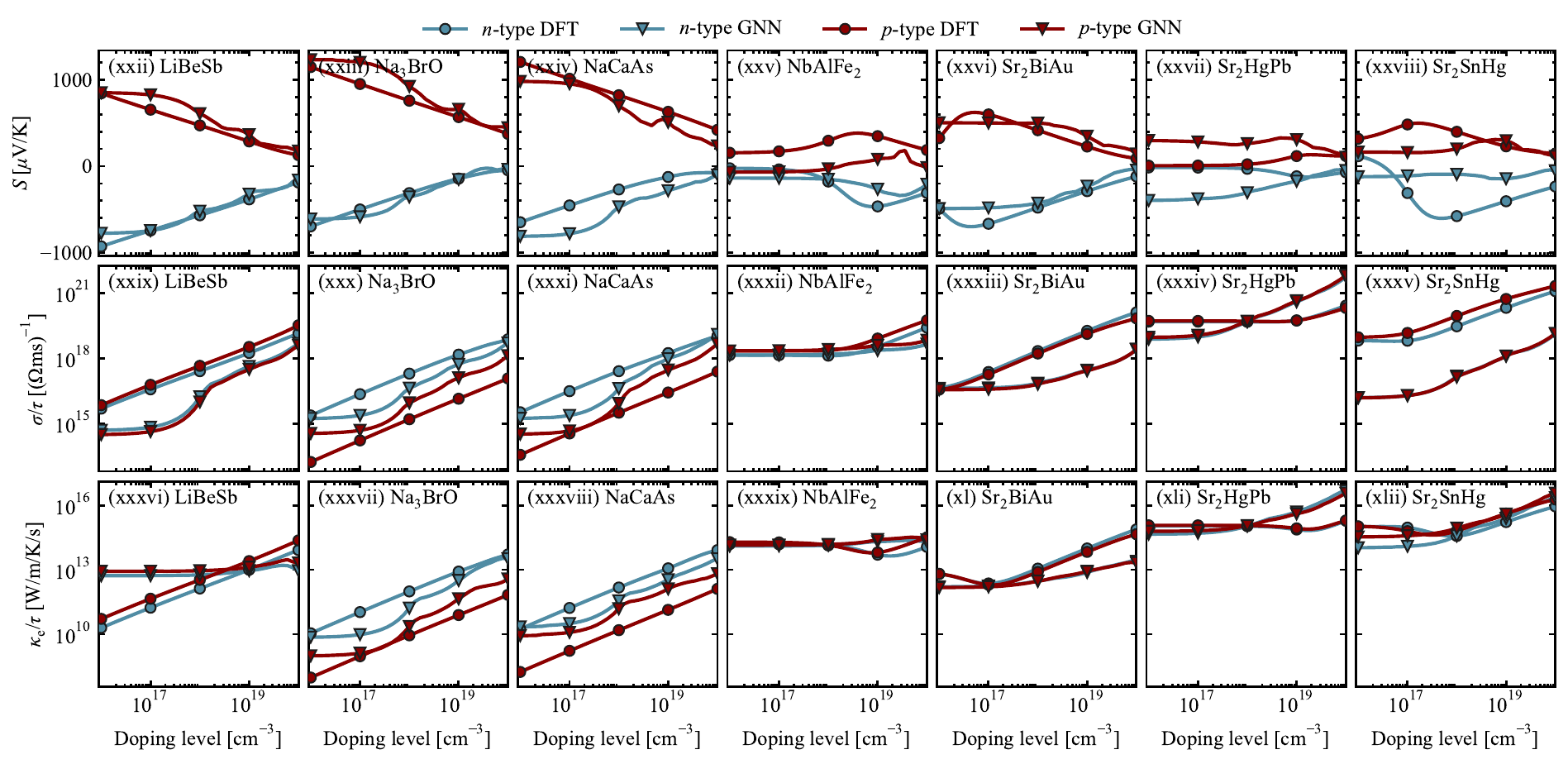}
    \includegraphics[width=.7\textwidth]{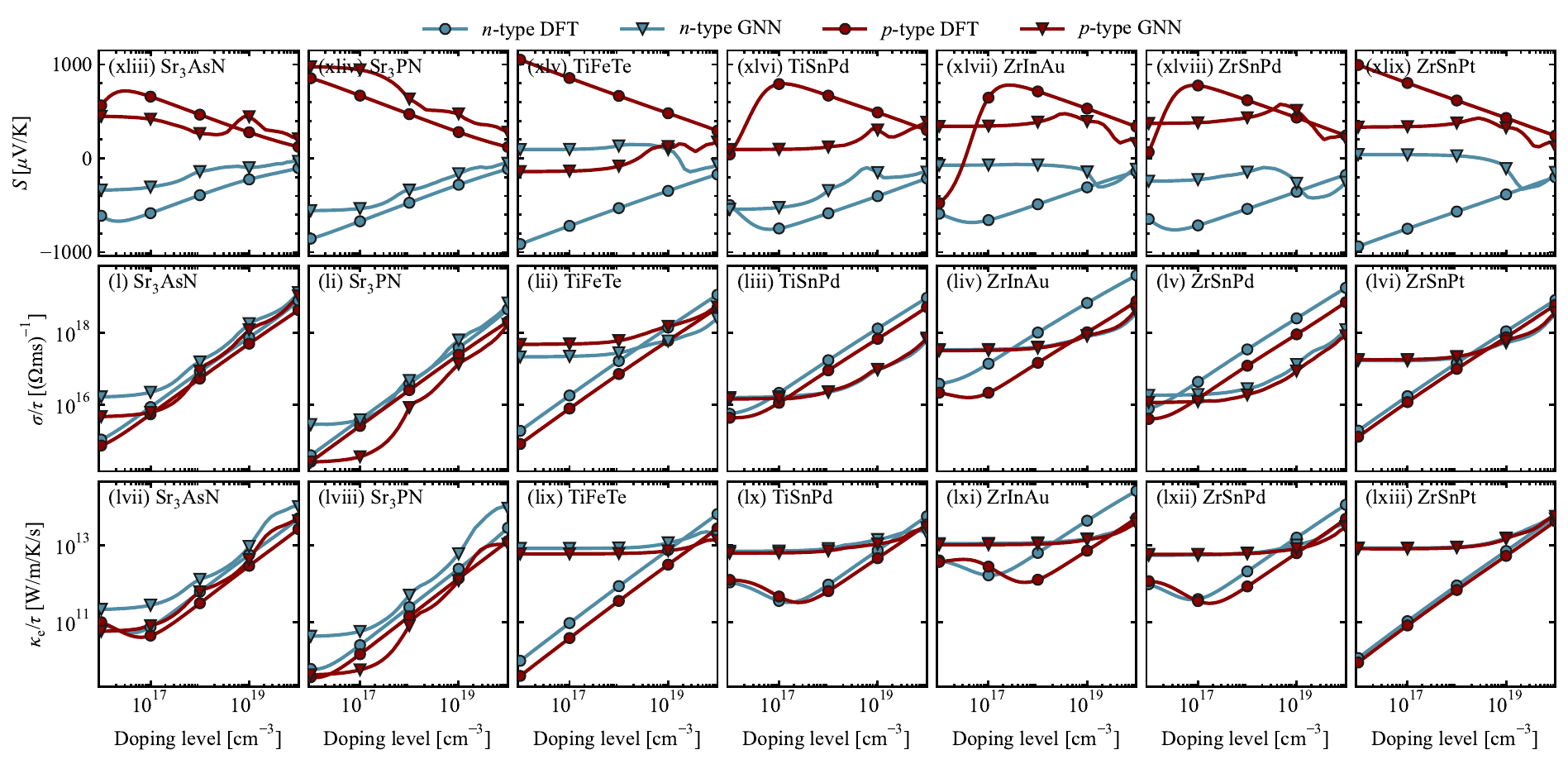}
    \caption{Carrier-concentration-dependent transport-property curves at 300~K for extended candidates derived from the Materials Project. Blue denotes $n$-type and red denotes $p$-type; circular markers represent the DFT-calculated values, while triangular markers represent the TECSA-GNN estimates.}
    \label{fig:extended-curves}
\end{figure}

For all candidates in the screening list for which reliable DFT calculations could be carried out, a total of 21 compounds, we performed complete calculations of the electronic transport coefficients with the doping type treated separately, and compared the results with the TECSA-GNN predictions. The corresponding results are shown in Fig.~\ref{fig:extended-curves}. This analysis provides a more comprehensive assessment of the actual predictive capability of TECSA-GNN. As shown in Fig.~\ref{fig:extended-curves}, TECSA-GNN exhibits excellent extrapolative performance in the vast majority of cases. At the same time, we also note two potential failure regimes. First, for compounds that may involve bipolar transport, such as the cases shown in Fig.~\ref{fig:extended-curves}(xlv)-(xlviii), the prediction of $S$ may become unreliable. Second, another potential failure regime occurs at low carrier concentrations for the prediction of $\left\{\sigma/\tau, \kappa_{\rm e}/\tau\right\}$. This issue is largely caused by the imbalance in the training-data distribution. As discussed above, weighted-loss training provides an effective way to alleviate this problem.

It should be noted that the results shown in Fig.~\ref{fig:extended-curves} were obtained by calculating the Onsager coefficients with the \texttt{AMSET} package~\cite{ganose2021efficient} based on the DFT electronic-structure parameters. In terms of both fidelity and computational cost, this treatment lies between CRTA and EPA, and therefore provides a more objective benchmark for assessing the performance of TECSA-GNN in a realistic high-throughput materials-screening workflow.


\end{document}